\documentclass[%
aip,
amsmath,amssymb,
reprint,%
]{revtex4-1}

\usepackage{graphicx}
\usepackage{dcolumn}
\usepackage{bm}

\usepackage[utf8]{inputenc}
\usepackage[T1]{fontenc}
\usepackage{mathptmx}

\usepackage[colorlinks=false,linkcolor=black]{hyperref}
\usepackage{xcolor, soul} 
\sethlcolor{green} 

\begin{document}
	
	\title{Harnessing vibrational resonance to identify and enhance input signals}
	
	\author{P. Ashokkumar} 
	\email{ak3phys@gmail.com}
	\affiliation{PG \& Research Department of Physics, Nehru Memorial College (Autonomous), Affiliated to Bharathidasan University, Puthanampatti, Tiruchirappalli - 621 007, India.}
	\affiliation{Department of Nonlinear Dynamics, School of Physics, Bharathidasan University, Tiruchirappalli - 620 024, India.}
	
	\author{R. Kabilan} 
	\email{kabilanrajagopal82@gmail.com}	
	\affiliation{PG \& Research Department of Physics, Nehru Memorial College (Autonomous), Affiliated to Bharathidasan University, Puthanampatti, Tiruchirappalli - 621 007, India.}
	
	\author{M. Sathish Aravindh} 
	\email{sathisharavindhm@gmail.com}
	\affiliation{Department of Nonlinear Dynamics, School of Physics, Bharathidasan University, Tiruchirappalli - 620 024, India.}
	
	\author{A. Venkatesan}
	\email{av.phys@gmail.com}
	\affiliation{PG \& Research Department of Physics, Nehru Memorial College (Autonomous), Affiliated to Bharathidasan University, Puthanampatti, Tiruchirappalli - 621 007, India.}

	\author{M. Lakshmanan}
	\email{lakshman.cnld@gmail.com}
	\affiliation{Department of Nonlinear Dynamics, School of Physics, Bharathidasan University, Tiruchirappalli - 620 024, India.}

	\begin{abstract}
	We report the occurrence of vibrational resonance (VR) and the underlying mechanism in a simple piecewise linear electronic circuit, namely the Murali-Lakshmanan-Chua (MLC) circuit, driven by an additional biharmonic signal with widely different frequency. When the amplitude of the high-frequency force is tuned, the resultant vibrational resonance is used to detect the low-frequency signal and also to enhance it into a high-frequency signal. Further, we also show that even when the low-frequency signal is changed from sine wave to square and sawtooth waves, vibrational resonance can be used to detect and enhance them into high-frequency signals. These behaviors, confirmed by experimental results, are illustrated with appropriate analytical and numerical solutions of the corresponding circuit equations describing the system. Finally, we also verify the signal detection in the above circuit even with the addition of noise.
	\end{abstract}
	\maketitle
		
	\begin{quotation}
	Vibrational resonance is a phenomenon that occurs in typical nonlinear systems where a weak low-frequency signal can be strengthened by a high-frequency signal, according to Landa and McClintock. The mechanism of vibrational resonance is the interaction of the two-frequency signals, which helps the quality of the weak low-frequency signal in the output to be enhanced. In this paper, we succeed in showing the enhancement of the quality of the weak low-frequency signal in the output of a driven Murali-Lakshmanan-Chua circuit by adding a high-frequency signal. The results of our investigation show that the different signals, namely sine, square, and sawtooth waves, can be identified and enhanced through vibrational resonance. These behaviors are confirmed by numerical and experimental studies. Also, we have verified the tolerance of the nature of the output against noise.
	\end{quotation}
		
	\section{Introduction}

	In a nonlinear system, the phenomenon of vibrational resonance (VR) takes place when weak periodic signals are amplified by the high periodic force present therein \cite{landa2000vibrational}. Due to the significance of their potential applications, high-frequency signal and VR phenomenon have been studied in different nonlinear dynamical systems \cite{rajasekar2016nonlinear}. There are many branches of science that use two-frequency signals \cite{roy2017vibrational}, including brain dynamics \cite{ge2020vibrational}, where bursting neurons may exhibit two different scales, and communication technology, where low-frequency signals are usually modulated by high-frequency signals\cite{landa2000vibrational,roy2021vibrational}, which encode the data \cite{palaniyandi2005estimation}.

	In particular, this phenomenon was first reported by Landa and McClintock \cite{landa2000vibrational} in a bistable system driven by a biharmonic force with widely different frequencies. Both computational and experimental reports of this VR effect in bistable electrical oscillators have been made \cite{baltanas2003experimental, bordet2012experimental}. Its potential applications are now being widely researched in a broad range of systems such as bistable systems \cite{lakshmanan2003chaos, rajasekar2016nonlinear}, multistable systems \cite{yang2010controlling, roy2017vibrational,roy2021vibrational}, excitable systems \cite{deng2014theoretical}, delayed dynamical systems \cite{yang2012vibrational,guo2020vibrational}, coupled neural oscillators \cite{deng2009effect,deng2010vibrational,wang2014vibrational, sarkar2019vibrational,ge2020vibrational}, and biological nonlinear systems \cite{ning2020vibrational}. In particular, the phenomenon of vibrational resonance is used in the implementation of logic gates \cite{venkatesh2016vibrational, gui2020enhanced, murali2021construction}, in the detection of faults in bearings \cite{xiao2019novel}, to detect a lower level input signal with a higher level amplitude of output \cite{chizhevsky2008vibrational, jia2018improving, xiao2020weak, pan2021study, yang2022research}, and also in the design of memory devices \cite{venkatesh2016vibrational, venkatesh2017implementation, aravindh2018strange, yao2022logical}.

	Recently, two of the present authors and Venkatesh have found that the quasiperiodically driven Murali-Lakshmanan-Chua (QPDMLC) system mimics dynamic logic gates and basic R-S flip-flops \cite{venkatesh2017design, sathish2020realisation}. In their study, they used two square waves to mimic the different logic behaviors and memory functions. In the recent literature, a few authors have applied vibrational resonance to investigate signal detection and enhancement. It was shown that a driven nano-electromechanical weak signal that has had its nonlinear resonance strengthened makes up a single-stable system \cite{chowdhury2020weak}. A high-frequency character signal has been formulated and fault detected using vibrational resonance in a stable state \cite{xiao2021adaptive}. Aperiodic and periodic weak signals were detected numerically using vibrational resonance \cite{chizhevsky2008vibrational, ren2017exploiting, ren2018generalized}. Generally, for vibrational resonance, one will use two forces, both of which are uniform \cite{rajasekar2016nonlinear}. Naturally, a question arises here as to whether one can change the low-frequency signal from sinusoidal to square or sawtooth force and fix the second force to be just sinusoidal and then predict the low-frequency signal (sine, square, and sawtooth) without altering the parameters or the second external force. We address this issue in this paper using a simple nonlinear circuit, both numerically and experimentally.

	A basic nonlinear circuit, namely Murali-Lakshmanan-Chua (MLC) circuit, characterized by its rich content and simplicity, has garnered significant attention in the literature regarding nonautonomous chaotic circuits, see Fig.\ref{fig2}(a) \cite{murali1994simplest, lakshmanan2003chaos}. It consists of an inductor, a capacitor, a resistor, and Chua's diode. As inductors are discrete in nature, they impose limitations on the ability to implement MLC circuits as integrated circuits (ICs). Conversely, in electronic circuits, the inductor is a less preferable circuit element. This is the result of a multitude of factors. For instance, inductors are comparatively less conventional than the remaining circuit components and necessitate individual preparation for the majority of applications. Furthermore, their spatial dimensions are more sizable, rendering them unsuitable for VLSI implementation unless the inductance is relatively negligible. Numerous inductorless implementations of chaotic circuits have been suggested as a consequence of this \cite{ccam2005inductorless, kilicc2007mixed, gunay2010mlc, swathy2014dynamics, gunay2017implementation, ashokkumar2021realization, fozin2022coexistence}.

	Compared to neural networks, state-controlled cellular neural networks (SC-CNNs) are a paradigm for parallel computation \cite{chua1988cellular}. A typical SC-CNN is composed of a vast quantity of identical, interconnected dynamical systems known as cells. The fundamental representation of these cells, or nodes is usually expressed in terms of coupled nonlinear ordinary differential equations. Narrow-band processing units are frequently denoted by the terms neurons or cells. Typically two CNN cells are utilized in the construction of the above MLC circuit, see Fig.\ref{fig2}(b), in addition to external forces such as biasing, noise, and sinusoidal force \cite{gunay2010mlc, swathy2014dynamics, ashokkumar2021realization}. Utilizing such an SC-CNN based MLC circuit to extend the analysis to the coupled system is extremely useful. 

	The theme of this paper is to study the underlying dynamical phenomenon of vibrational resonance in the presence of two different forces in a piecewise linear non-autonomous system, and in particular in the Murali-Lakshmanan-Chua (MLC) system using the State Controlled - Cellular Neural Networks (SC-CNNs). We have also carried out appropriate numerical studies on the SC-CNN MLC circuit when additional low and high-frequency signals are present. Also, in the present study, we point out that, in contrast to the earlier investigation in Ref.\cite{venkatesh2016vibrational}, for the appropriate choice of system parameters, one can identify VR with a maximal response over a wide range of high-frequency amplitudes, which will enable assured applications in signal detection and enhancement. Particularly the low input signals are in the form of sinusoidal, square wave, or sawtooth types. The obtained numerical results for the detection and enhancement of low-frequency signals are compared and confirmed with the corresponding experimental results. Finally, we demonstrate a robust principle to detect the low-intensity signal via vibrational resonance even in the presence of additional Gaussian white noise.

	The structure of this paper is as follows: In Section \ref{sec2}, we provide an overview of the notion of vibrational resonance in the SC-CNN cell structure of the nonlinear, non-autonomous system. In Sections \ref{sec3}, \ref{sec4}  and \ref{sec5}, we discuss the numerical and experimental realization, respectively, of response amplitude in detecting the low-frequency signal and its enhancement for different input signals. In Section \ref{sec6}, the impact of noise is examined. We conclude by summarizing our findings in Section \ref{sec7}. Also, we discuss the analytical evaluation of the response amplitude in Appendix \ref{sec_a}.

	\section{Vibrational resonance in state controlled CNN based nonlinear non-autonomous system}
	\label{sec2}

	\begin{figure}[]
	\centering
	\includegraphics[width=0.9\columnwidth]{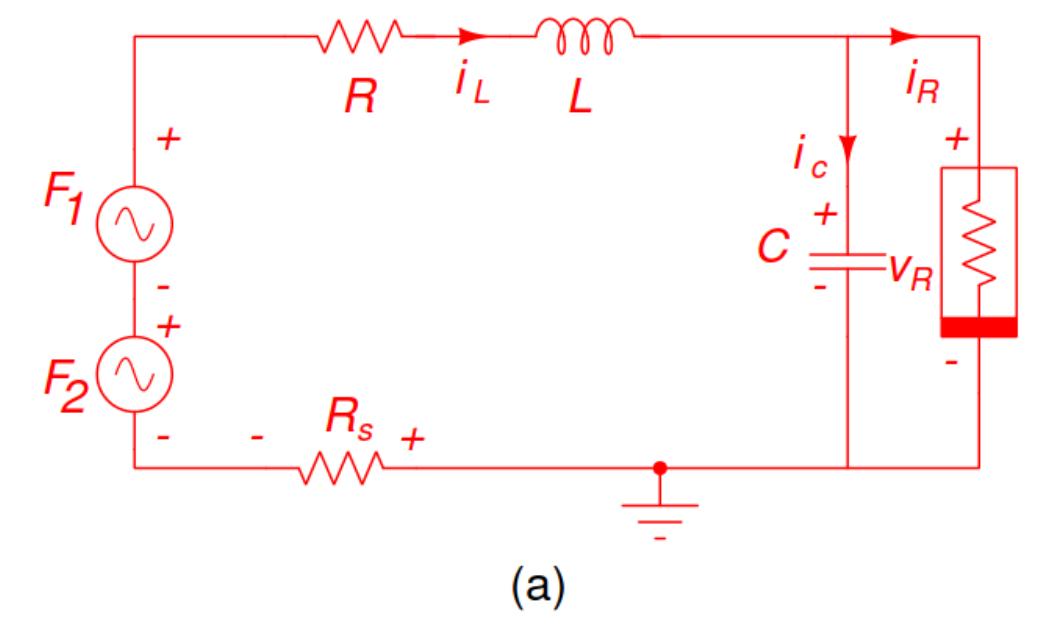}
	\includegraphics[width=0.9\columnwidth]{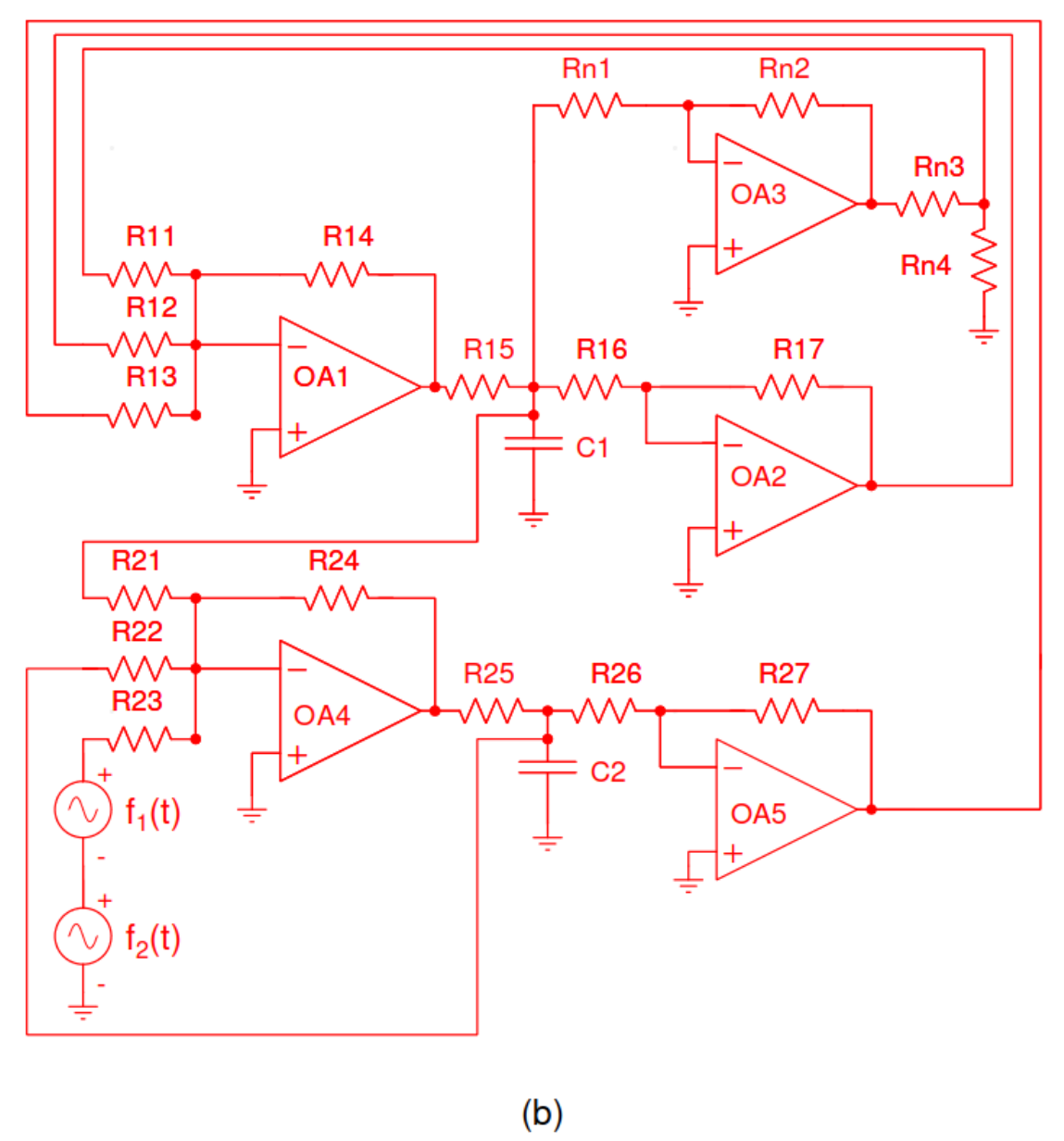}
	\caption{Panel (a) and (b) shows the schematic diagram for MLC circuit and SC-CNN based MLC circuit.}
	\label{fig2}
	\end{figure}

	The concept of Cellular Neural Network (CNN) was first developed by Chua and Yang in 1988 \cite{chua1988cellular}. It is an n-dimensional array of resistors, capacitors, OP-AMPs, and other analog circuit components, but without any inductors. The underlying CNN is built from a large number of interconnected dynamical systems. 

	CNN is a reasonably basic circuit that may be easily implemented experimentally using appropriate electronic circuit elements. These circuits are powerful tools for the emulation of complex dynamics in nonlinear systems. In this case, the local output and voltage variables of the CNN cells are exchanged with one another. This generalization uses the analog components of CNN and it is known as the State Controlled - CNN (SC-CNN). Many chaotic circuits designed and implemented in terms of SC-CNN have been documented in the literature \cite{gunay2010mlc, ashokkumar2021realization, arena1995simplified,  gunay2017implementation, swathy2013experimental, luo2016dynamics,  li2022vibrational}. The significant advantages of these CNN circuits are fourfold, that is (1) they have no inductors, (2) their only circuitry is RC based, (3) they are parallelly connected, and (4) they consume less power. Thus the SC-CNN circuits are realized with less number of hardware and are easily implemented in VLSI design \cite{manganaro2012cellular}. The present study illustrates that, after adding an additional sine wave signal to this circuit, the resultant output exhibits similar and inverted enhanced sine wave signals. Further, this phenomenon is also illustrated for as other signals such as square wave and sawtooth wave.

	The single forced SC-CNN MLC circuit has been well studied at the numerical, experimental, and analytical levels in ref.\cite{gunay2010mlc}. The standard MLC circuit [see Fig.\ref{fig2}(a)] consists of a nonlinear resistor that has the three-segment piecewise characteristics of Chua’s diode, a linear resistor, a linear inductor, and a linear capacitor with a sinusoidal voltage source \cite{murali1994simplest,lakshmanan2003chaos}.

	It is a well-established fact \cite{murali1994simplest, lakshmanan2003chaos} that the normalized form of the MLC circuit equation with additional sinusoidal force connected in series (Fig.\ref{fig2}(a)) can be written as
	\begin{eqnarray}
	\dot x&=& y - h(x), \nonumber \\
	\dot y&=& -\beta (1+\nu)y - \beta x + f_{1} \sin (\omega_{1} t)+f_{2} \sin (\omega_{2} t),~~~~~~
	\label{equ2}
	\end{eqnarray}
	where the piecewise linear function h(x) is given as
	\begin{equation}
	h(x) =  \left \{
	\begin{array} {ll} 
		bx + (a - b),   & x > 1,     \\
		ax,             & |x| \le 1,     \\
		bx - (a - b),   & x < -1.     
	\end{array}
	\right.
	\label{equ3}
	\end{equation}

	The relationship between the various circuit variables and circuit parameters and the above dynamical variables and parameters can be obtained from Refs.\cite{murali1994simplest, lakshmanan1996chaos, lakshmanan2003chaos}. Earlier studies on the dimensionless version of the circuit with a single external periodic force for the chosen parametric values $a=-1.02, \: b=-0.55, \: \gamma = 0.015, \: \beta = 1.0$ and $\omega_{1} = 0.75$ is readily available in the literature. In these studies, the quantity $f_{1}$, which is the  amplitude  the first periodic force, was varied to identify different bifurcation structures. It has been shown that this dynamical system exhibits chaos via different routes, including the period-doubling route, intermittency route, and strange non-chaotic attractor (SNA) route, among others \cite{lakshmanan1996chaos, lakshmanan2003chaos}. 

	The dynamic model of the circuit (Fig.\ref{fig2}(b)) of two generalized CNN cells is in accordance with the following coupled state equations, 
	\begin{eqnarray}
	\dot{x}_{1} & = &-x_{1}+a_{11}y_{1}+a_{12}y_{2}+\sum_{k=1}^{2}s_{1k}x_{k}+i_{1}, \nonumber \\
	\dot{x}_{2} & = &-x_{2}+a_{21}y_{ 1}+a_{2}y_{2}+\sum_{k=1}^{2}s_{2k}x_{k}+i_{2},
	\label{equ5}
	\end{eqnarray}
	where $ x_{1} $ and $ x_{2} $ are state variables, and $ y_{1} $ and $ y_{2} $ are the corresponding outputs. The MLC circuit equation defined by Eq.(\ref{equ2}) can be derived from Eq.(\ref{equ5}), by assuming $ x=x_{1} $, $ y=x_{2} $, $ a_{1}=b-a $, $ a_{12} = a_{21}=a_{2}=0$, $ s_{11}=1-b $, $ s_{12}=1 $, $ s_{21}=-\beta $, $ s_{22}=1-\beta(1+\nu) $, $ i_{1}=0 $ and $i_{2} = f_{1} \sin (\omega_{1} t)+f_{2} \sin (\omega_{2} t) $. 

	\begin{figure}[]
	\centering
	\includegraphics[width=1.0\linewidth]{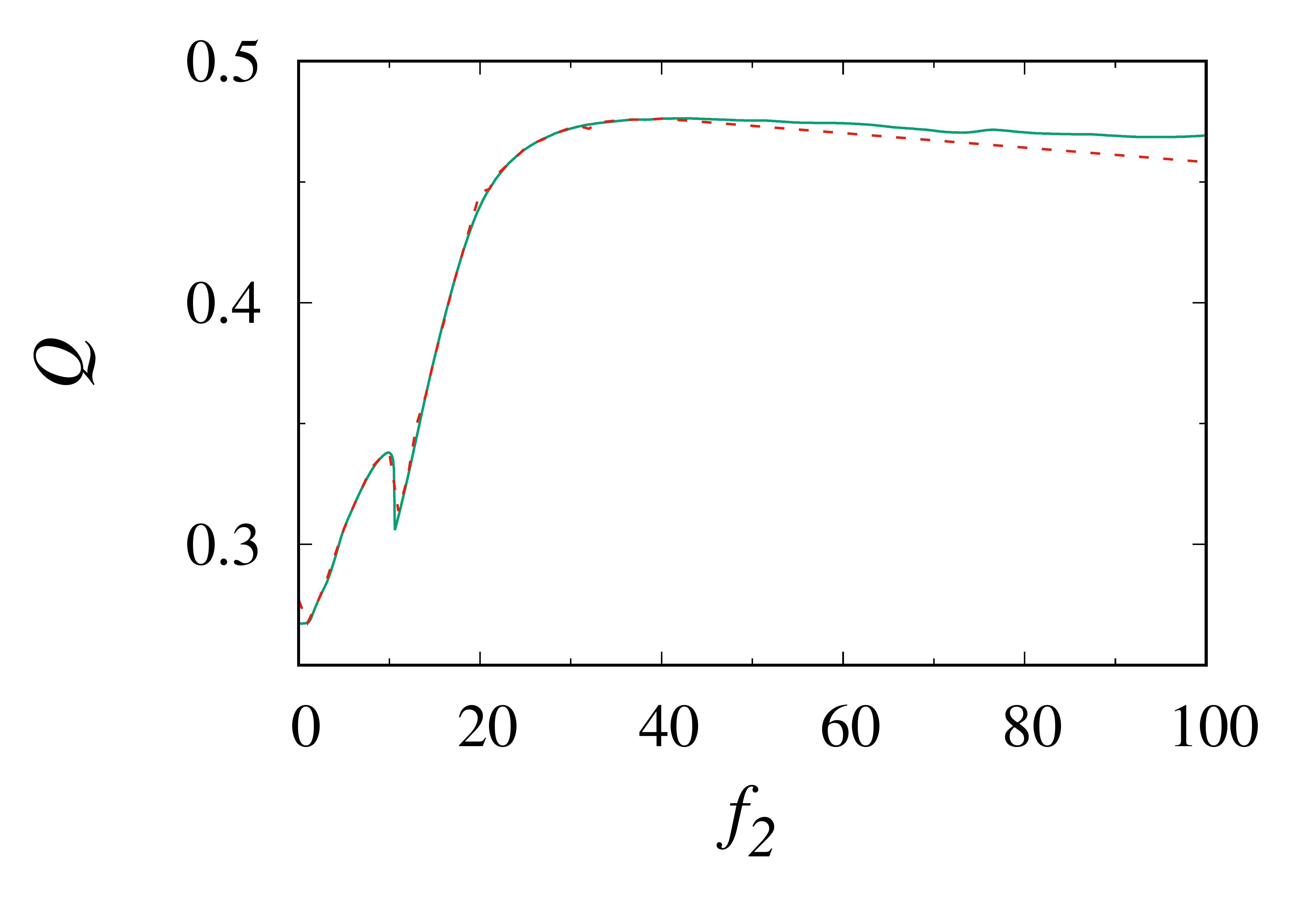}
	\caption{The dependence of response amplitude Q with $ f_{2} $. The solid line and dashed line represent the numerical and analytical response amplitudes, respectively,  for a fixed set of values of parameters with $\omega_1 = 0.75$, $\omega_2 = 3.5$, and the forcing parameter $f_1 = 0.25$.}
	\label{fig4}	
	\end{figure}

	\begin{figure}[!ht]
	\centering
	\includegraphics[width=1.0\linewidth]{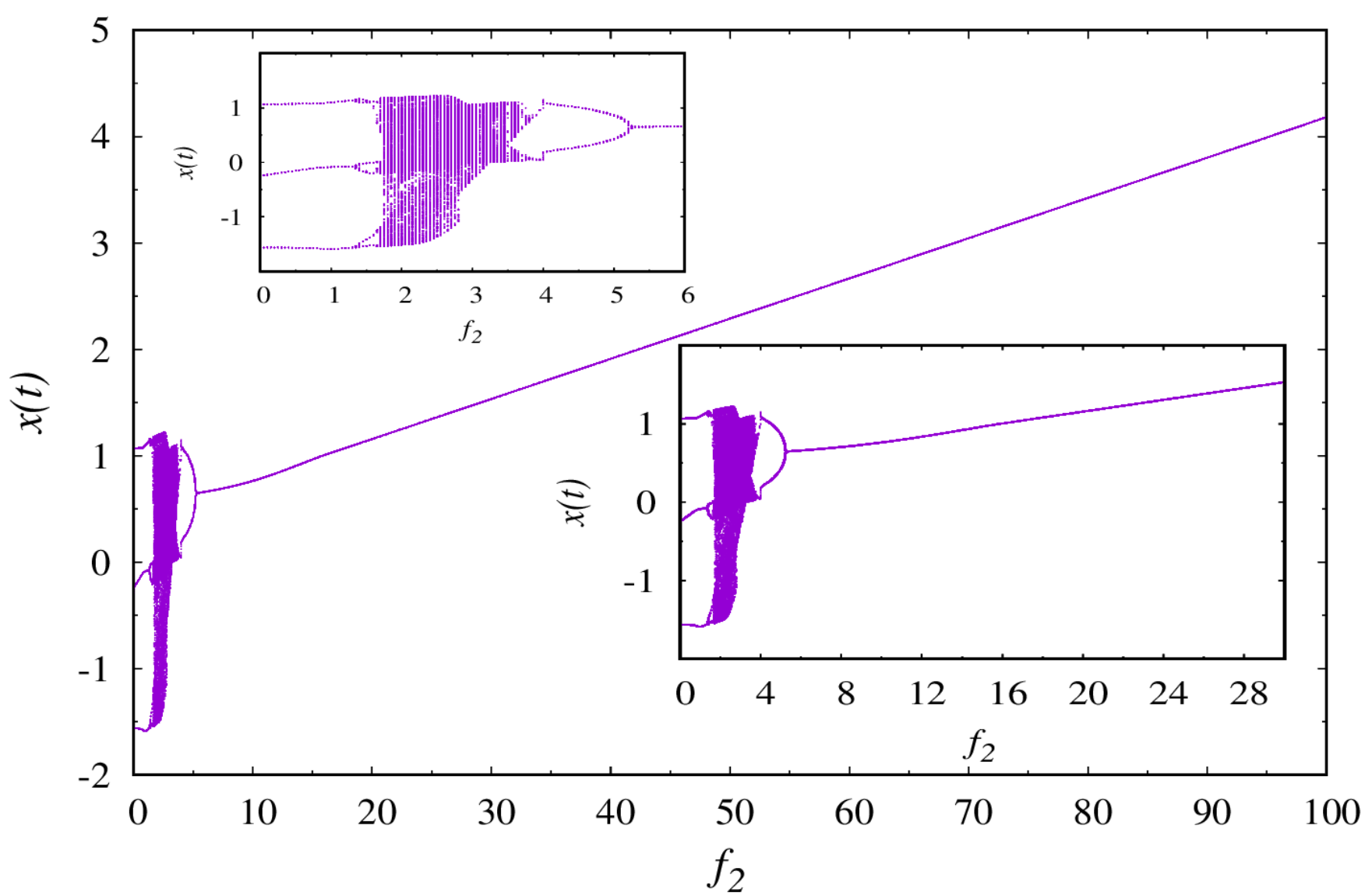}
	\caption{Bifurcation diagram of $ f_{2} $ vs $ x(t) $. The sub-figure shows the enlarged version of external force $ f_{2} $ for a fixed set of values of parameters with $\omega_1 = 0.75$, $\omega_2 = 3.5$, and the forcing parameter $f_1 = 0.25$.}
	\label{fig13}	
	\end{figure}

	\begin{figure}[!ht]
	\centering
	\includegraphics[width=1.0\linewidth]{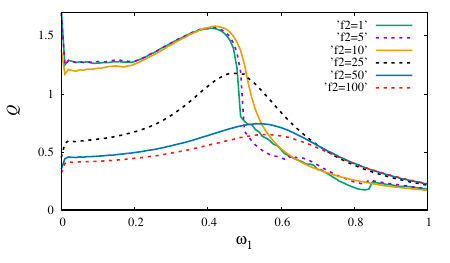}
	\caption{The dependence of response amplitude Q with $ \omega_{1} $ for different values of $ f_{2} $. Different colors indicate different values of the second forcing parameter $ f_{2} $. The remaining parameters are fixed with $\omega_2 = 3.5$, and the first forcing parameter $f_1 = 0.25$.}
	\label{fig14}	
	\end{figure}

	\begin{figure}[!ht]
	\centering
	\includegraphics[width=1.0\linewidth]{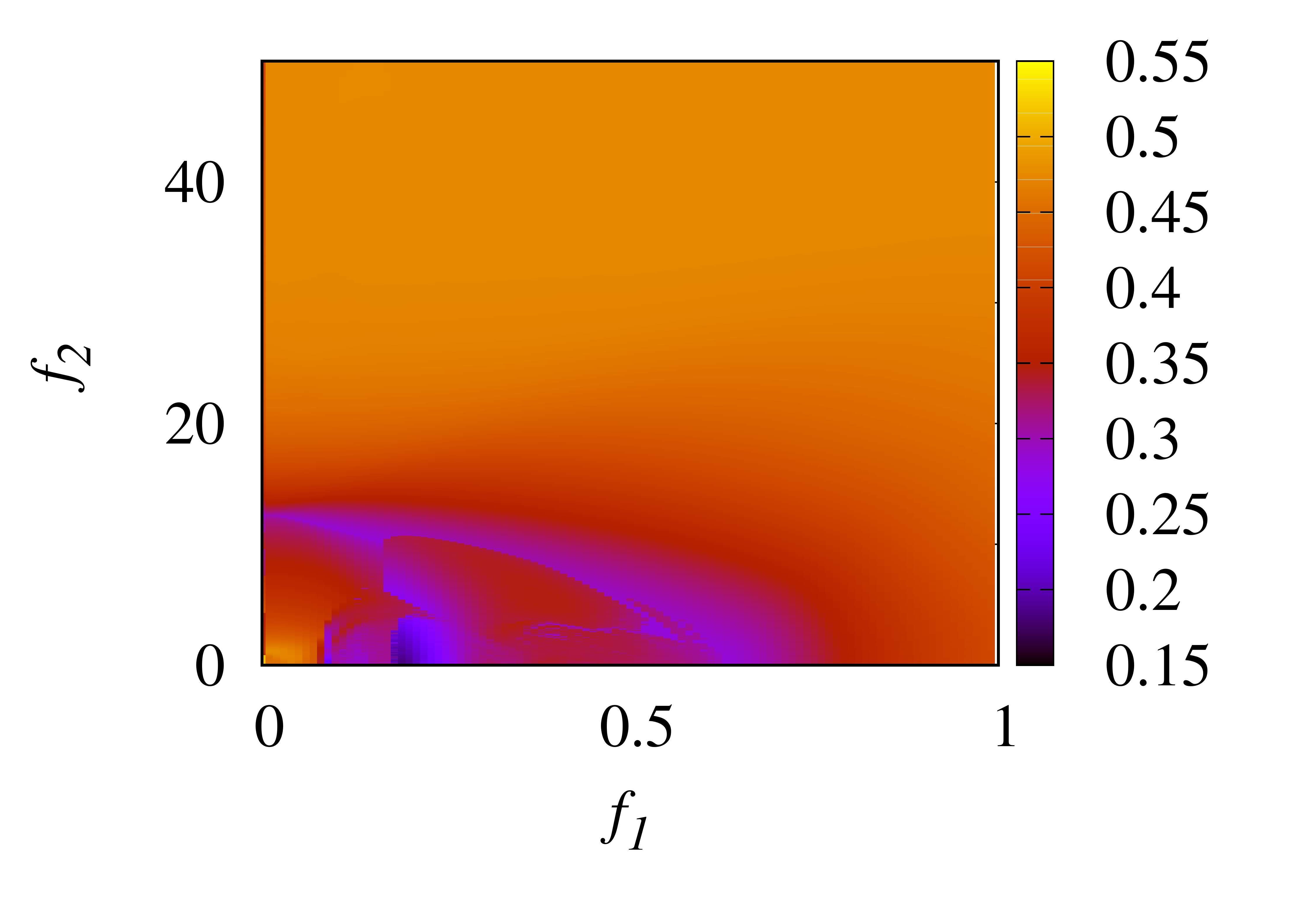}
	\caption{The maximum response amplitude Q depicted in a two-parameter phase diagram for the amplitude of the first force $ f_{1} $ vs second force $ f_2 $ for fixed parameters $\omega_1 = 0.75$ and $\omega_2 = 3.5$.}
	\label{fig5}	
	\end{figure}
	Consequently, from Eq.(\ref{equ5}), the SC-CNN based MLC circuit model (Fig.\ref{fig2}(b)) is organized as below : 
	\begin{eqnarray}
	\dot{x}_{1}&=&-x_{1}+a_{1}y_{1}+s_{11}x_{1}+s_{12}x_{2}, \nonumber \\
	\dot{x}_{2}&=&-x_{2}+s_{21}x_{1}+s_{22}x_{2}+f_{1} \sin (\omega_{1} t)+f_{2} \sin (\omega_{2} t),~ 
	\label{equ8}
	\end{eqnarray}
	where $y_1$ is the piecewise linear function,
	\begin{equation}
	y_1 = f(x_1) = 0.5(\mid x_1 + 1 \mid - \mid x_1-1 \mid).
	\label{equ8a}
	\end{equation}
	Here, the external forces $ f_{1} sin (\omega_{1} t) $  and $ f_{2} sin (\omega_{2} t) $ correspond to low and high level frequencies, respectively, where the amplitude $ f_{1}=0.25 $ is fixed while $ f_{2} $ is varied. Further, the following are typically the rescaled parameter values: $a_1=0.47; ~s_{11}=1.550; ~s_{12}=1.0; ~s_{21}=-1.0; ~s_{22}=-0.015; ~\omega_1=0.75;~\omega_2=3.5;$ and $f_1=0.25$. For suitable choices of the control parameter value $f_2$, the system exhibits a rich variety of dynamics. In the present paper, we investigate the effect of vibrational resonance, detect the existence of a low input frequency signal in the corresponding response, and also obtain the enhancement of it to a high-frequency signal in this circuit.

	\section{Analytical and Numerical evaluation of response amplitude}
	\label{sec3}

	\begin{figure*}[t]
	\centering 
	\includegraphics[width=1.0\columnwidth]{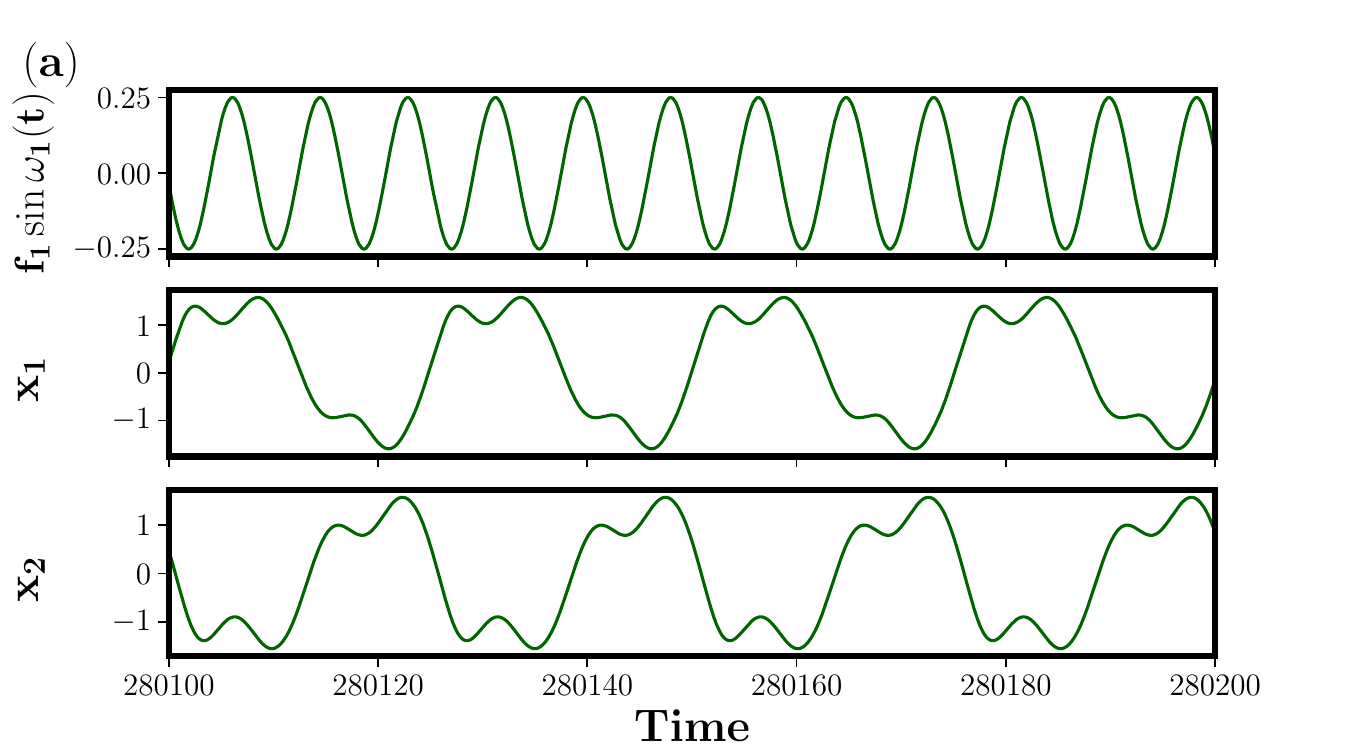}
	\includegraphics[width=1.0\columnwidth]{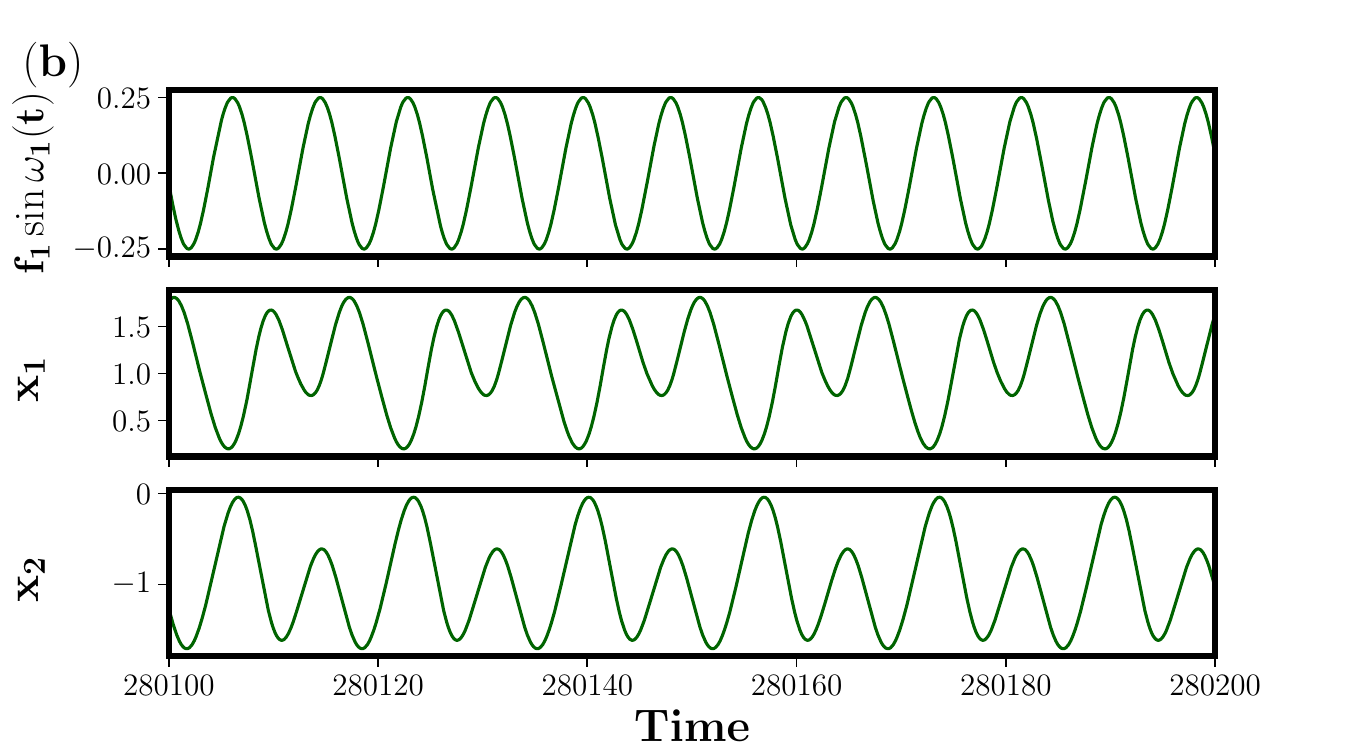}
	\includegraphics[width=1.0\columnwidth]{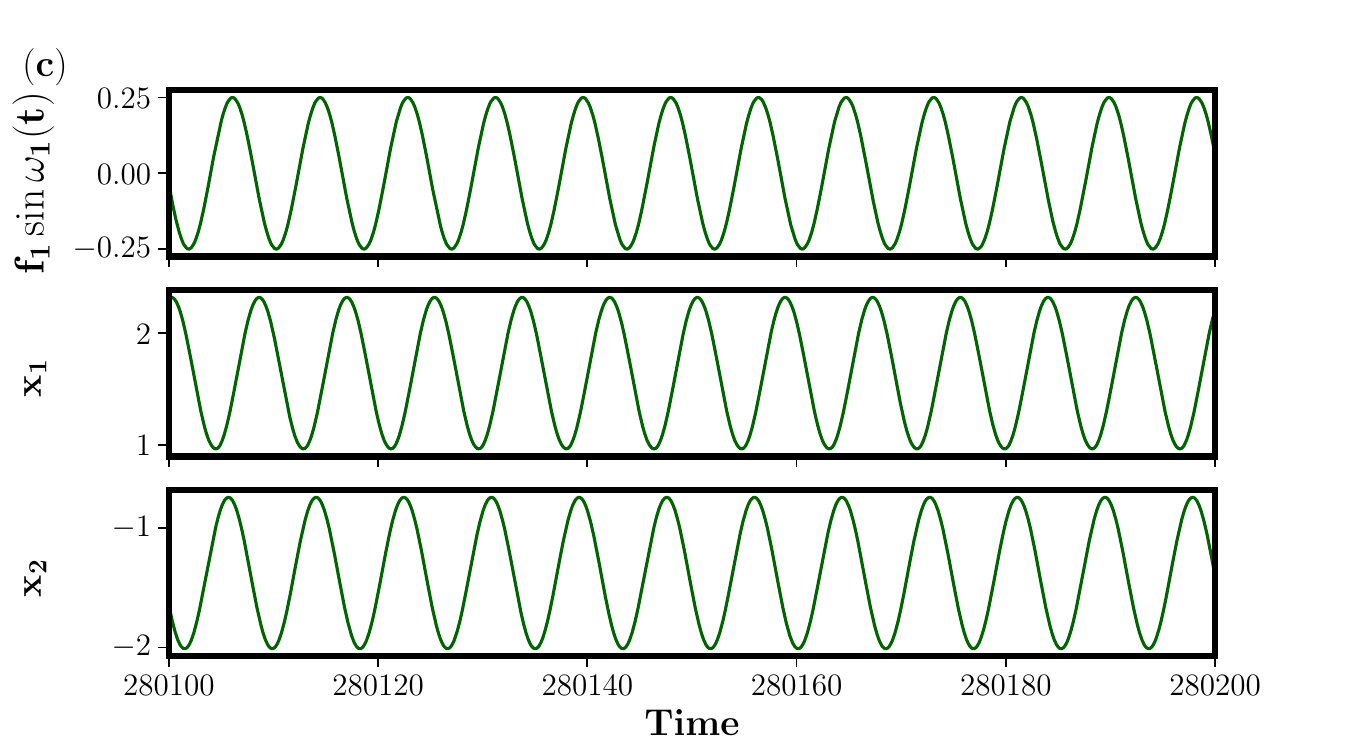}
	\includegraphics[width=1.0\columnwidth]{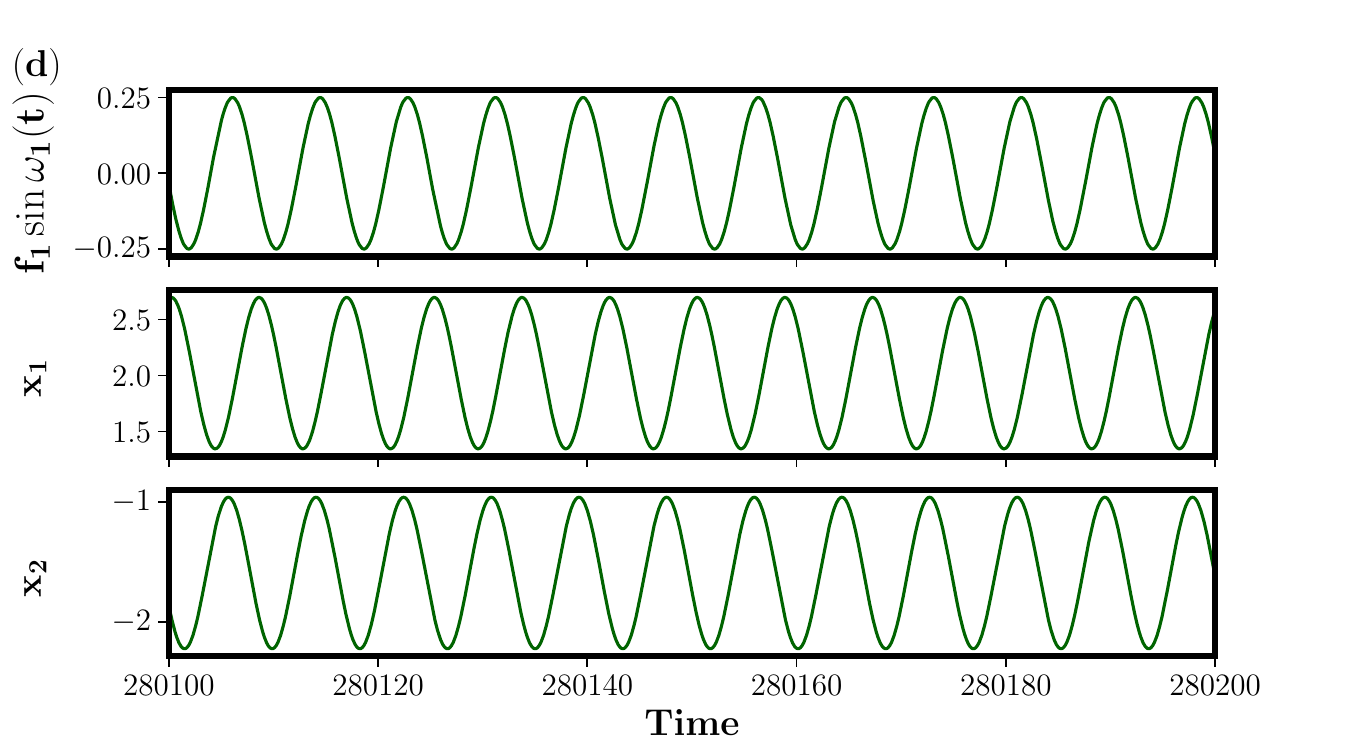}
	\caption{Realization of the signal detection in the numerical simulation : Panels (a)-(d)  correspond to different values of $ f_{2} $, $ f_{2}=1.0,5.0,15.0 $ and $ 25.0 $, respectively. Every panel is having three sub-figures, namely the low amplitude signal $ f_{1} $, and the resultant outputs $ x_{1} $ and $ x_{2} $. }
	\label{fig7}
	\end{figure*}
	The analytical expression for the response function can be explicitly obtained using the exact solution of the dynamical equation of the MLC circuit equation (\ref{equ2})-(\ref{equ3}) in the three different piecewise regions of the Chua's diode. This is carried out in Appendix \ref{sec_a}. Consequently, we obtain 
	\begin{equation}
	Q_{ana} = \frac{\sqrt{(\sum Q_s^{0,\pm})^2 + (\sum Q_c^{0,\pm})^2}}{f_1},
	\label{equ21}	
	\end{equation}
	where $ Q_s^{0,\pm} $, $ Q_s^{0,\pm} $ are given in Eqs.(\ref{a14}) and (\ref{a15}) in Appendix \ref{sec_a}. To obtain the corresponding response amplitude from numerical analysis, we proceed as follows.

	After solving numerically Eq.(\ref{equ8})  and using the results in the following expression for the response function, after discarding the transients, we obtain \cite{landa2000vibrational}
	\begin{equation}
	Q_{num} = \frac{\sqrt{\rule{0pt}{2ex} Q_{s}^2+Q_{c}^2}}{f_{1}}, 
	\label{equ9}
	\end{equation}
	over a range of values of the forcing strength of high-frequency driving force ($f_2$). Here the values of the quantities $Q_{s}$ and $Q_{c}$ are computed from the Fourier spectrum of the time series of the output signal $x(t)$ as
	\begin{eqnarray}
	Q_{c} = \frac{2}{nT} \int_{0}^{nT} x(t) \cos (\omega_{1} t) dt, \nonumber \\
	Q_{s} = \frac{2}{nT} \int_{0}^{nT} x(t) \sin (\omega_{1} t) dt,
	\label{equ10}
	\end{eqnarray}
	where n is an integer. The response curves of $ Q $ versus selected parameters for a range of system parameters are numerically calculated. The system (\ref{equ8}) is numerically integrated with Runge-Kutta fourth order algorithm with step size $\Delta t = 0.01$. 

	\begin{figure}[]
	\centering
	\includegraphics[width=0.8\columnwidth]{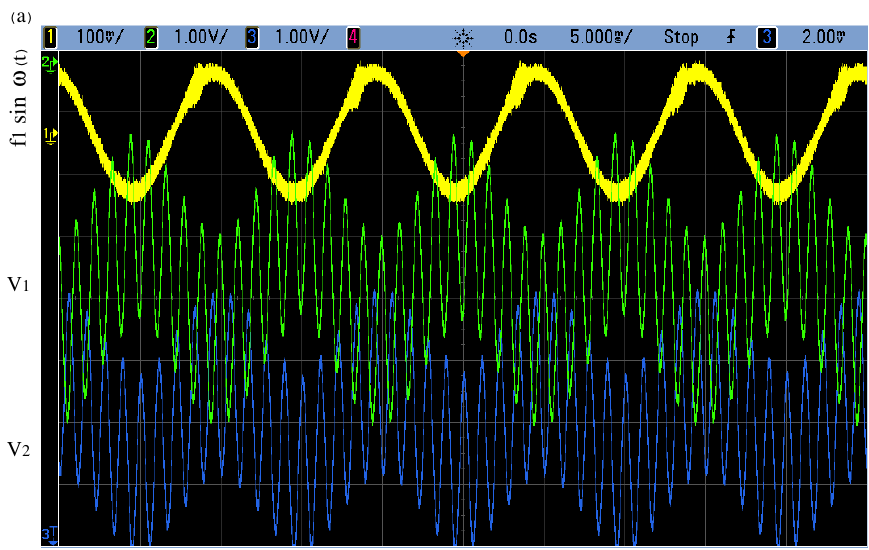}
	\includegraphics[width=0.8\columnwidth]{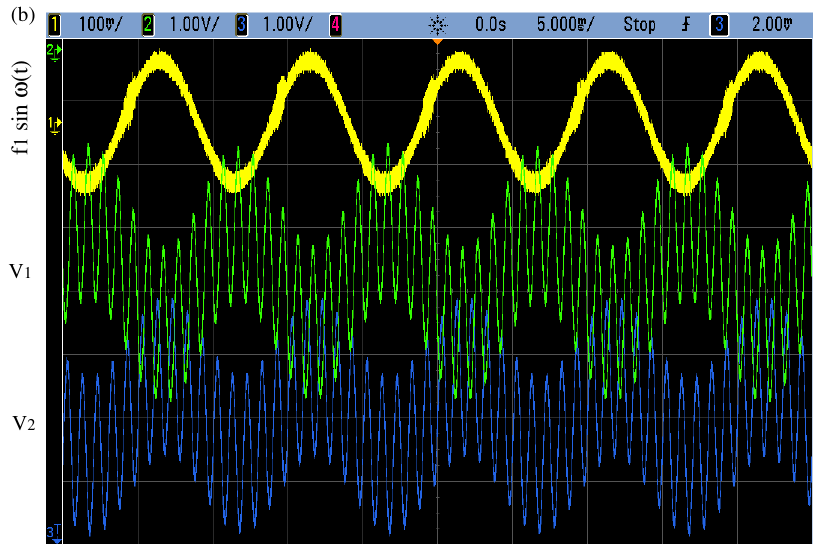}
	\includegraphics[width=0.8\columnwidth]{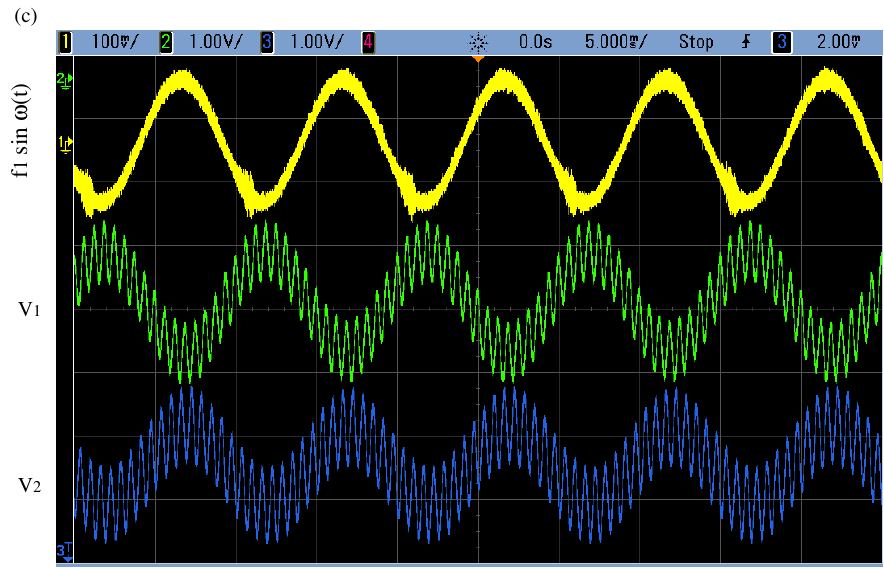}
	\includegraphics[width=0.8\columnwidth]{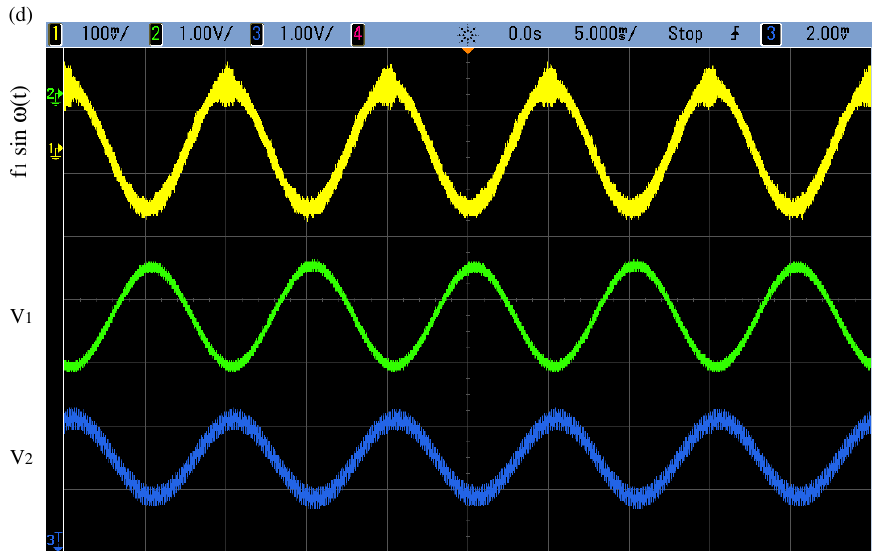}
	\caption{Realization of the signal detection in experimental SC-CNN based MLC electronic circuit : Panels (a)-(d) correspond to different values of $ f_{2} $, $ f_{2}=100$ mV, $300$ mV, $700$ mV , and $ 1.0 $~V, respectively. Every panel is having three sub-figures namely the low amplitude signal $ f_{1} $, and the experimental outputs $ v_{1} $ and $ v_{2} $.}	
	\label{fig3}
	\end{figure}

	The analytically and numerically calculated response amplitudes from Eq.(\ref{equ21}) and Eq.(\ref{equ9}) are shown in Fig.\ref{fig4}. The solid and dashed lines represent the numerical and analytical response amplitudes, respectively. To obtain these curves, we fix the system parameter values as $\omega_1 = 0.75$, $\omega_2 = 3.5$ and $f_1 =  0.25$. The remaining parameter, $f_{2}$, is selected in such a way that it promotes the emergence of VR. It is obvious from Fig.\ref{fig4} that as the high-frequency force $ f_{2} $ increases, the response amplitude slowly increases and reaches a first maximum value when $f_{2}=10.0$. After this, there is a sudden drop in the response amplitude, and then it again increases until the response amplitude reaches its maximum value around $ f_{2} = 25.0 $. Any further increase in $ f_{2} $ does not lead to any notable change in the value of $ Q $. It is also instructive to look at the nature of the corresponding dynamics, which can be identified from the structure of the associated bifurcation diagram. Fig.\ref{fig13} exhibits the bifurcation diagram for different values of the high forcing parameter $ f_{2} $. Initially, for low $ f_{2} $ the system exhibits period-3 oscillations ($ f_{2} \in [0,1.30] $); on further increase in $ f_{2} $, the system exhibits period doubling phenomenon in the range of $ f_{2} \in [1.30,1.65] $ and exhibits chaotic behavior for $ f_{2} \in [1.65,3.66] $. After this beyond $ f_{2}=3.66 $ the system shows reverse period doubling phenomenon for the range of $ f_{2} \in [3.66,5.3] $. After the forcing parameter $ f_{2} $ reaches the value $ f_{2}=5.3 $, the system exhibits period-1 oscillations. Further, by increasing the forcing parameter $ f_{2}>5.3 $, the system continuously exhibits period-1 oscillations behavior. Also, with the increase of $ f_{2} $, the response Q reaches a maximum value when $f_{2}=10.0$, after which one observes a sudden drop in the response amplitude (see Fig.\ref{fig4}), and then it again increases as $ f_{2} $ increases until $ f_{2} $ reaches the value around $ f_{2}=25.0 $ beyond which saturation arises. Here again the system exhibits limit cycle oscillations in this range of $ f_{2} \in [5.3,100] $ \cite{murali1994simplest,lakshmanan2003chaos}. The sub-figure Figs.\ref{fig13} show the enlarged version as a function of the external forcing parameter $ f_{2} $, $ f_{2} \in [0,30] $.

	Further in Fig.\ref{fig14}, we present the response amplitude against the first forcing frequency $\omega_{1}$ for different values of the forcing parameter $f_{2}$. Different values of $ f_{2} $ are indicated by different colors in Fig.\ref{fig14}. Initially, at a low-frequency of $f_{2} $, we obtain a high response at a low value of $\omega_{1}$. On increasing the value of $ f_{2} $, the maximum of the response amplitude shifts to larger value of $\omega_{1}$. For all the values of $ f_{2} $, the response curves exhibit a saturation at higher values of $\omega_{1}$.

	Fig.\ref{fig5} is a 3-dimensional plot that depicts the numerically computed response as a function of the first forcing strength $ f_{1} $ of low-frequency signal and the second forcing strength $ f_{2} $ of high-frequency. It is clearly demonstrated in Fig.\ref{fig5} that the response amplitude ‘$ Q $’ is constant for the higher value of the second forcing strength $ f_{2} $. Increasing the low-frequency signal $ f_{1} \in [0,1] $, the response amplitude attains the maximum value for a low value of the second forcing signal $ f_{2} $. In Fig.\ref{fig7} every panel indicates the three signals, namely the low input signal $ f_{1} $ and the outputs $ x_{1} $ and $ x_{2} $. Let us consider first the low input signal for $f_1=0.25$ and the second forcing signal $f_2=1.0$. The corresponding outputs $x_1$ and $x_2$ are shown in Fig.\ref{fig7}(a). The two outputs $x_1$ and $x_2$ randomly oscillate and do not match with the low input signal $f_1$. On further increasing the $f_2$ value to $f_2=5.0$, the outputs $x_1$ and $x_2$ slowly approach the actual input behavior [see Fig.\ref{fig7}(b)]. At $f_2=15.0$, we find the outputs $x_1$ and $x_2$ approach exactly the sinusoidal behavior where the output $x_2$ matches with the low input signal $f_1$, and the other output $x_1$ inversely matches it. Also, the output signals get enhanced in comparison with the input signal $f_1$ [see Fig.\ref{fig7}(c)]. Then, on further increase in the $f_2$ value to $f_2=25.0$ and beyond, that is in the range $ f_{2} \in [25,200] $, the outputs $x_1$ and $x_2$ accurately predict the frequency of the input signal $f_1$ continuously in the form of inverse and similar signals, respectively [see Fig.\ref{fig7}(d)]. Also, the response outputs are enhanced in comparison with the low input signal $ f_{1} $. 

	\section{Experimental realization}
	\label{sec4}

	Using experimental realization, in the following, we verify the results obtained by numerical simulation discussed in the earlier sections. Now we consider the biharmonic forced SC-CNN based Murali-Lakshmanan-Chua circuit given in Fig.\ref{fig2}(b) with the circuit cell components $ R_{11} $ = 207 $K\Omega $, $ R_{12} $ = 66 $K\Omega $, $R_{13}$ = 100 $K\Omega $,  $ R_{14} $ = 100 $K\Omega $, $ R_{15} $ = 1 $K\Omega $, $ R_{16} $ = 100 $K\Omega $, $ R_{17} $ = 100 $K\Omega $, $ R_{n1} $ = 220 $K\Omega $, $ R_{n2} $ = 3 $M\Omega $, $ R_{n3} $ = 180 $ K\Omega $, $ R_{n4} $ = 16 $ K\Omega $, $ R_{21} $ = 100 $ K\Omega $, $ R_{22} $ = 6666.6 $ K\Omega $, $ R_{23} $ = 100 $ K\Omega $, $ R_{24} $ = 100 $ K\Omega $, $ R_{25} $ = 1 $ K\Omega $, $ R_{26} $ = 100 $ K\Omega $, $ R_{27} $ = 100 $ K\Omega $,  $ R_{31}-R_{38} $ = 10 $ K\Omega $, $ C_1 $ = 10 $ nF $, $ C_2 $ = 10 $ nF $ and active element $IC741$ type voltage OP-AMPs with $\pm12~V$ supply voltages. Experimental results were obtained using Agilent function generators $ (33220A) $ and Agilent digital storage oscilloscopes $ (DSO~7014B) $. The changes in the dynamics of the circuit under the effect of input streams are obtained by measuring the voltages $v_1$ and $v_2$ across the capacitors $C_1$ and $C_2$, respectively. Further, we vary the second forcing parameter $f_2$ from 100 mV to 1 V. In this case, Fig.\ref{fig3}(a) shows the low input signal $ f_{1} $ and two output signals $ v_{1} $ and $ v_{2} $. The first input $f_1$ is fixed as 100 mV and the second input $f_2$ is also given the value 100 mV. The corresponding outputs of the system signals do not predict the low-frequency input signals $f_1$ (fig.\ref{fig3}(a)). As $f_2$ is increased to 300 mV (fig.\ref{fig3}(b)) the lower input is not observed in the output signal. On further increase to the value $f_2=700$ mV (see fig.\ref{fig3}c), the output signals are able to replicate the low input signals $ f_{1} $. Even further increase of $ f_{2} $ to $f_2 = 1.0$ V (see Fig.\ref{fig3}d), the lower input signal is predicted from the output signal $ v_{2} $ (and its inverse from $ v_{1} $) and also the amplitude gets enhanced from $ 100$ mV to $ 1 $ V (see the panel at the top of every figure in Fig.\ref{fig3}). On further increase in the value of $f_2$, the output signals are able to detect the lower input signal continuously.

	\section{Analyzing the behavior of different low-frequency input signals $ f_{1} $}
	\label{sec5}

	\begin{figure}[h!]
	\centering 
	\includegraphics[width=1.0\columnwidth]{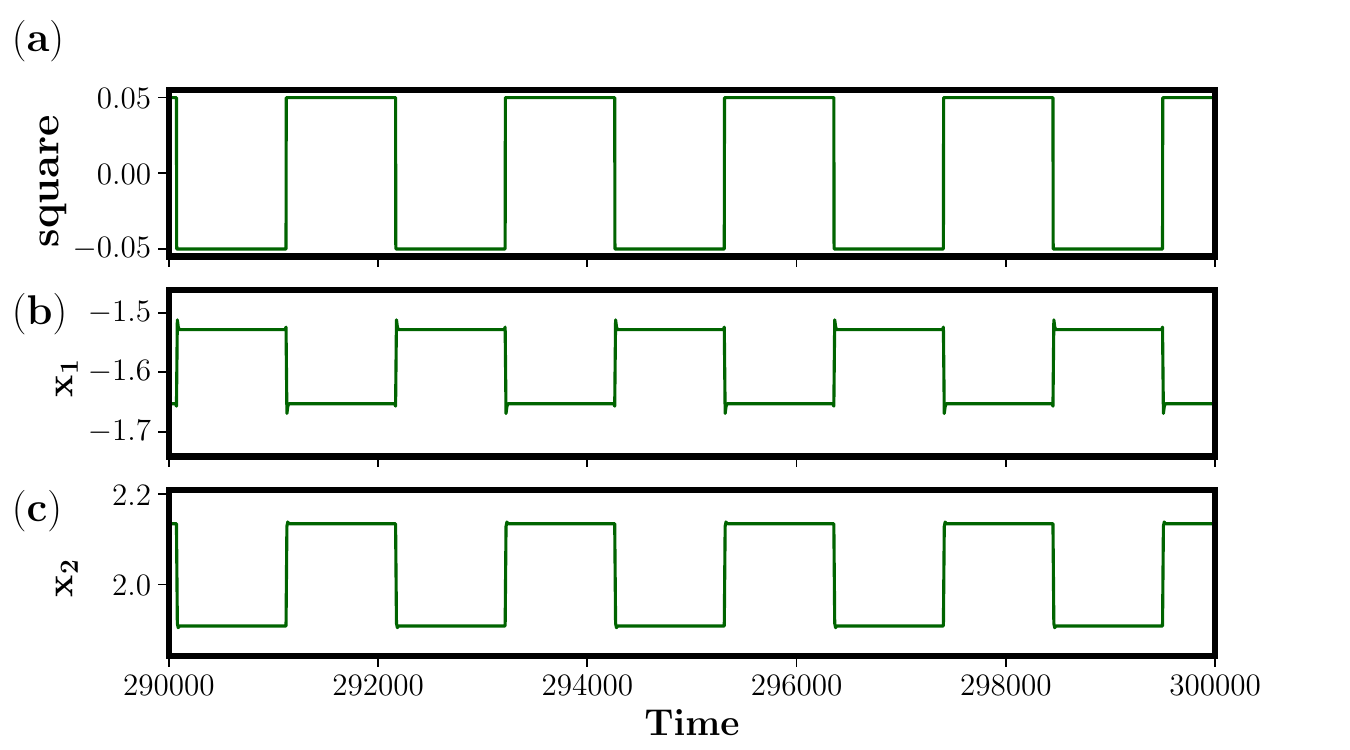}
	\caption{Realization of the signal detection in the numerical simulation: Panels (a)-(c)  correspond to low amplitude square wave, and the resultant outputs $ x_{1} $ and $ x_{2} $ for a fixed value of $ f_{2}=25.0 $.}
	\label{fig11a}
	\end{figure}

	\begin{figure}[h!]
	\centering 
	\includegraphics[width=1.0\columnwidth]{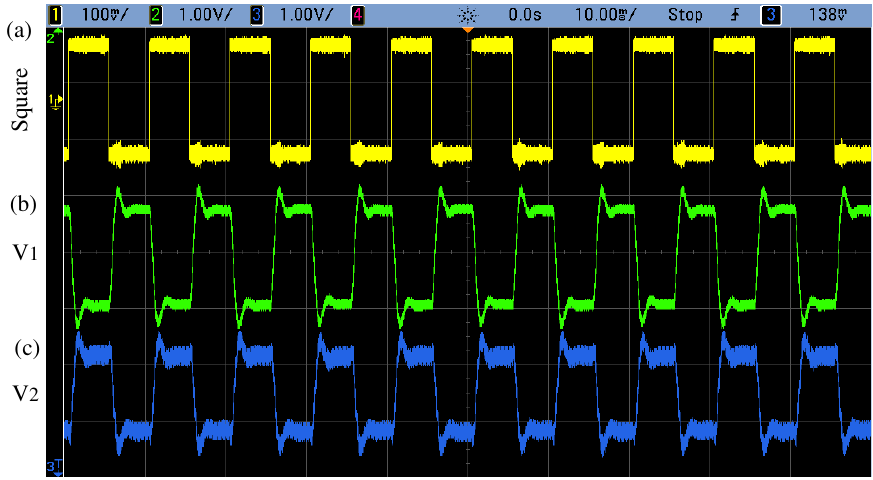}
	\caption{Realization of the signal detection in the experimental realization: Panels (a)-(c)  correspond to low input of square wave, and the resultant experimental outputs $ v_{1} $ and $ v_{2} $.}
	\label{fig11b}
	\end{figure}

	\begin{figure}[h!]
	\centering 
	\includegraphics[width=1.0\columnwidth]{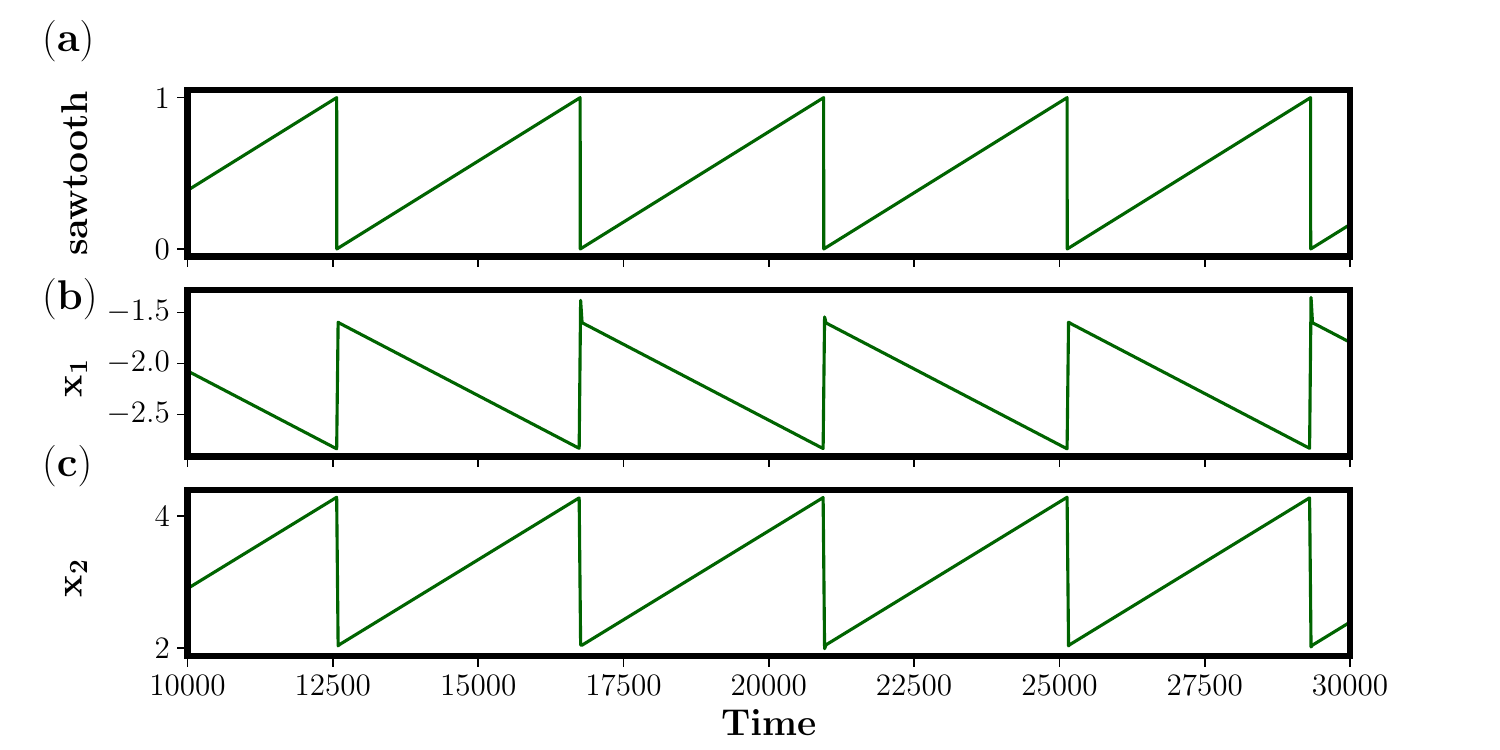}
	\caption{Realization of the signal detection in the numerical simulation: Panels (a)-(c)  correspond to low amplitude sawtooth wave, and the resultant outputs $ x_{1} $ and $ x_{2} $ for a fixed value of $ f_{2}=25.0 $.}
	\label{fig12a}
	\end{figure}

	\begin{figure}[h!]
	\centering 
	\includegraphics[width=1.0\columnwidth]{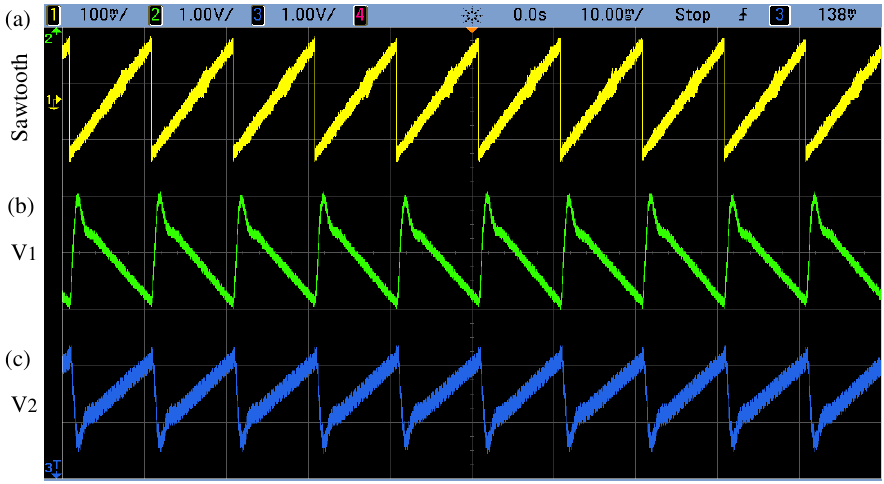}
	\caption{Realization of the signal detection in the experimental realization: Panels (a)-(c)  correspond to low input of sawtooth wave, and the resultant experimental outputs $ v_{1} $ and $ v_{2} $.}
	\label{fig12b}
	\end{figure}

	Next, we pose an interesting question: is it possible to detect other types of wave profiles like square, sawtooth, etc. waves which are different from sinusoidal ones? The answer is yes if one changes the low-frequency signal from a sine to a square or sawtooth wave, as discussed below. For this purpose, we fix the high-frequency signal $ f_2 $ as a sine wave and change the low-frequency signal $ f_1 $ from a sine wave to a square or sawtooth wave. Figs.\ref{fig11a} and \ref{fig11b} show the corresponding numerical and experimental outputs of the square wave. Also, Figs.\ref{fig12a} and \ref{fig12b} show the corresponding numerical and experimental outputs of the sawtooth wave. In particular, Figs.\ref{fig11a} and \ref{fig12a} indicate the three panels (a)-(c), namely the low input signal square/sawtooth wave and the corresponding numerical outputs $ x_{1} $ and $ x_{2} $. The experimental outputs Figs.\ref{fig11b} and \ref{fig12b} indicate the three panels ($ a)-(c) $, namely the low input signal square/sawtooth wave and the corresponding resultant outputs $ v_{1} $ and $ v_{2} $. The corresponding numerical and experimental outputs  $x_1/v_{1}$ and $x_2/v_{2}$ are shown in Figs.\ref{fig11a}/\ref{fig12a} and \ref{fig11b}/\ref{fig12b}. These outputs $x_1/v_{1}$ and $x_2/v_{2}$ approach exactly the square and sawtooth waveforms and accurately predict the frequency of the square and sawtooth wave input signals continuously in the form of inverse and similar signals, respectively [see Figs.\ref{fig11a}/\ref{fig12a} and \ref{fig11b}/\ref{fig12b}]. Also, the response outputs are enhanced in comparison with the low input signal. Thus, we confirm using different input signals like square/sawtooth wave signals that we are able to detect and enhance the output signals through numerical and experimental realizations.

\begin{figure*}[]
\centering 
\includegraphics[width=1.0\columnwidth]{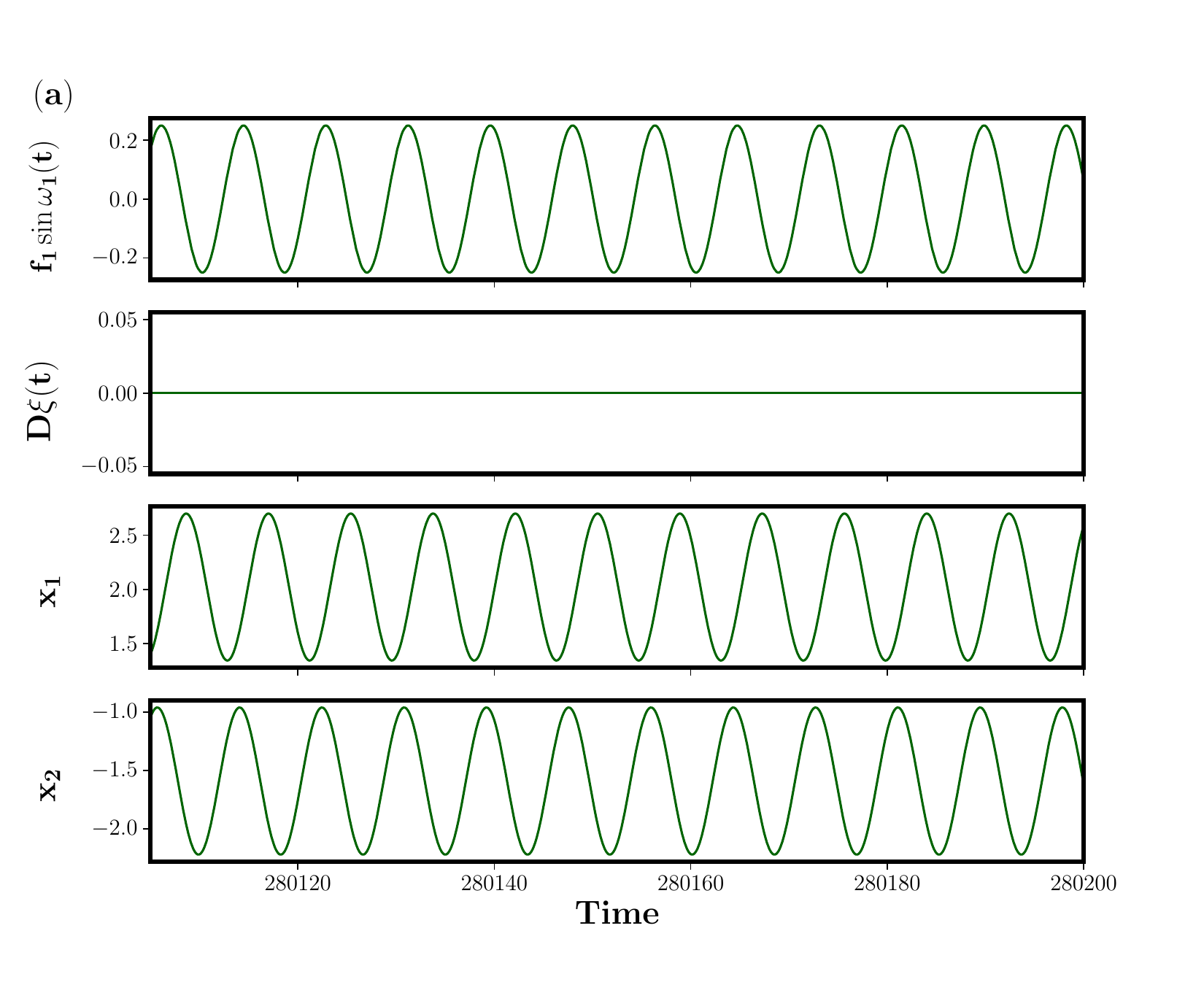} 
\includegraphics[width=1.0\columnwidth]{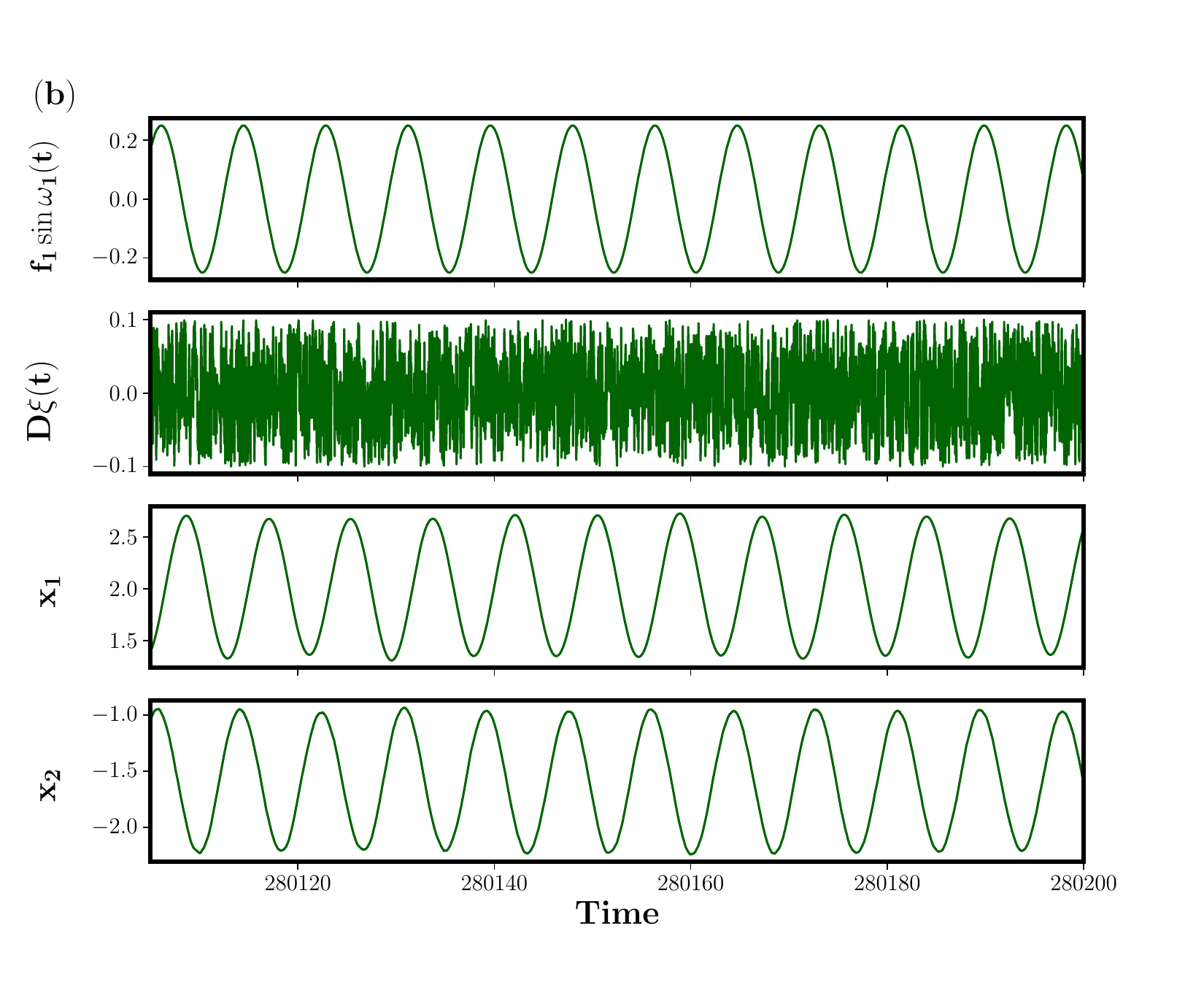} 
\includegraphics[width=1.0\columnwidth]{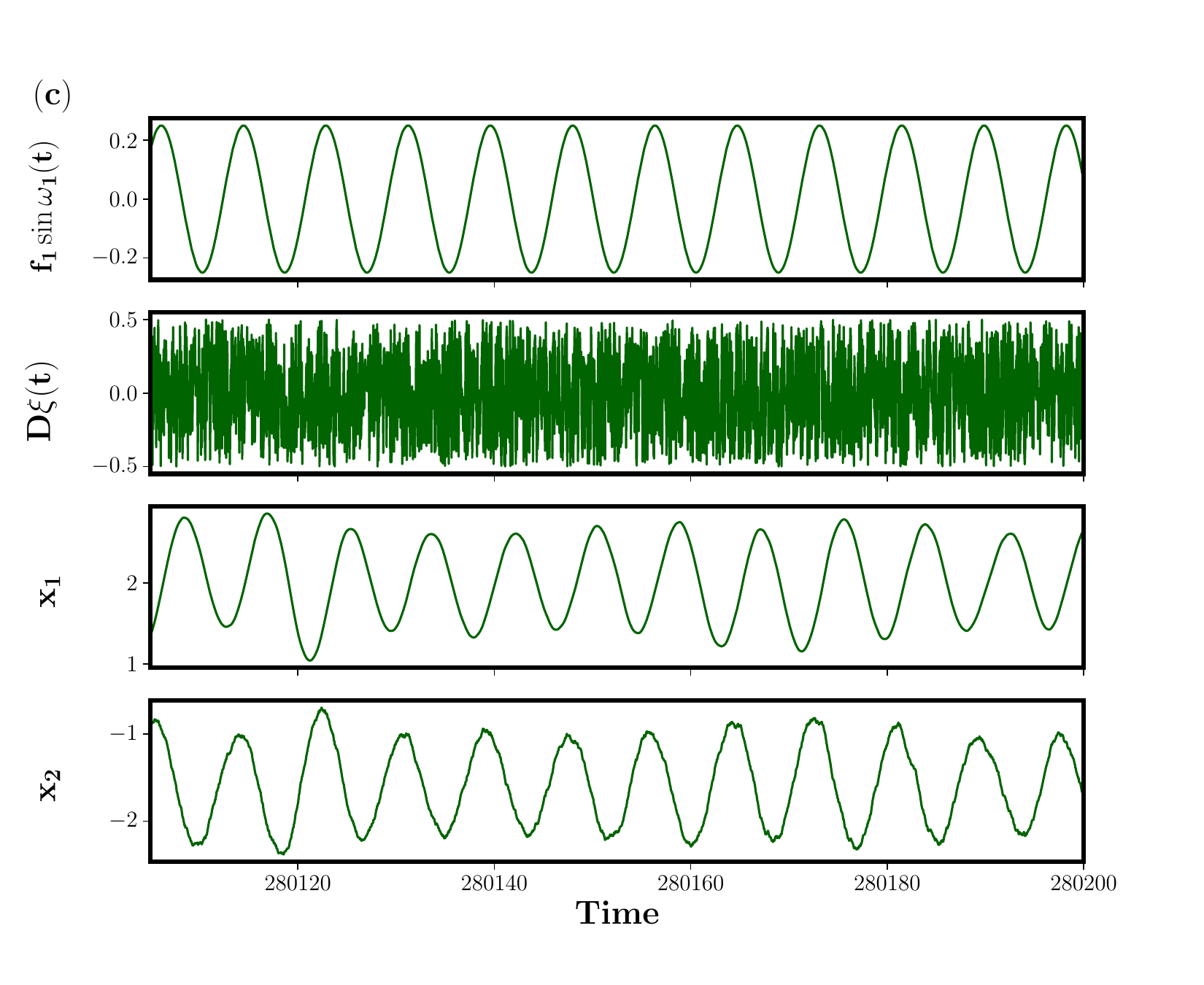} 
\includegraphics[width=1.0\columnwidth]{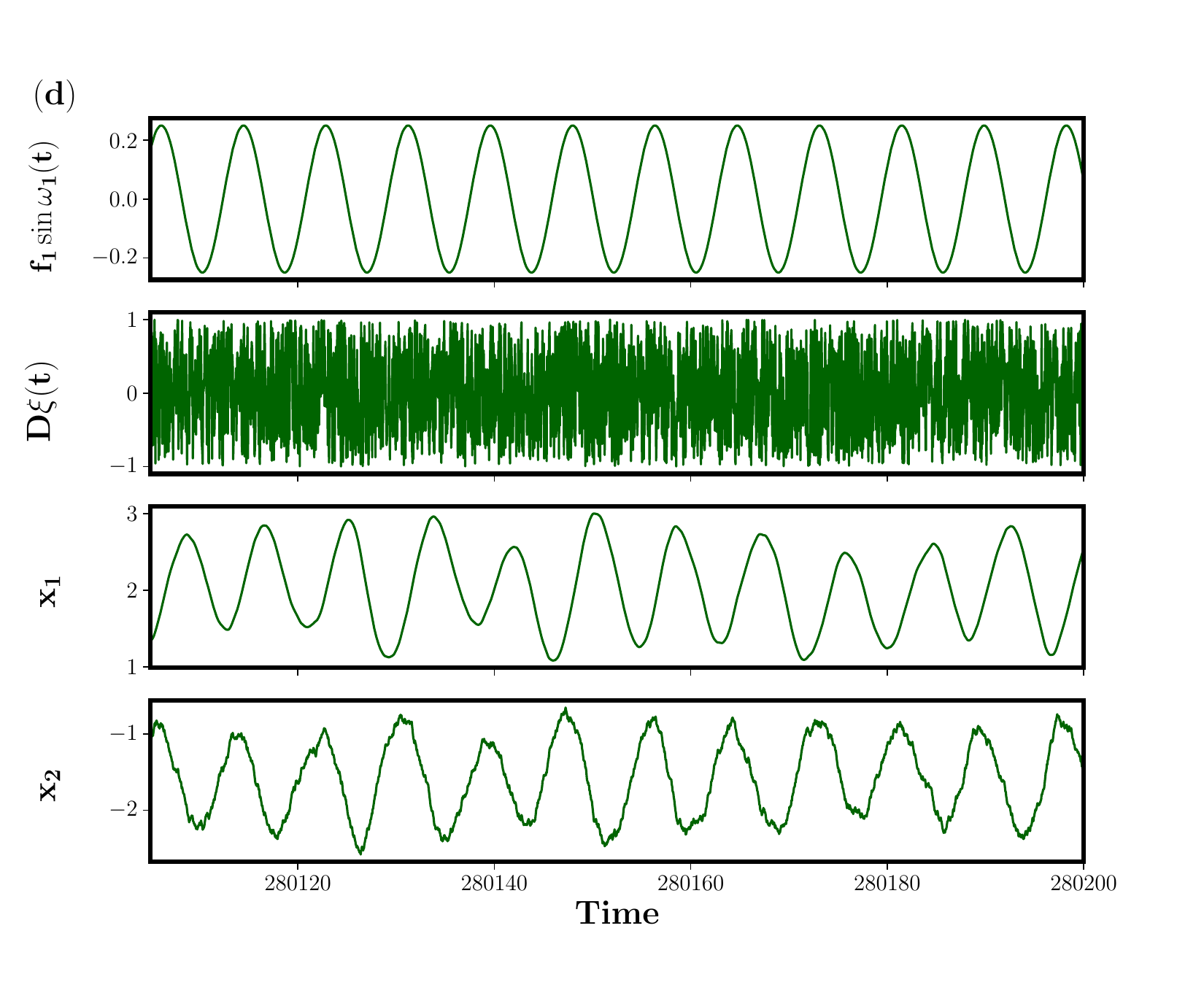}
\caption{Realization of the signal detection in numerical simulation: Panels (a)-(d) corresponds to different values of $ D $. Panel (a) represents $ D=0.0 $, panel (b) represents $ D=0.1 $, panel (c) represents $ D=0.5 $ and panel (d) represents $ D=1.0 $ with fixed forcing parameter $ f_{1}=0.25 $ and $ f_{2}=25.0 $. Every panel is having four subfigures namely  low amplitude $ f_{1} \sin \omega_{1} t $, Gaussian white noise $D \xi (t)$, and the resultant outputs $ x_{1} $ and $ x_{2} $.}
\label{fig8}
\end{figure*}

	\section{Effect of noise}
	\label{sec6}

	At this point, a question that may naturally arise is whether or not the system exhibits the same kind of structure after the inclusion of additional noise. Now we reexpress Eq.(\ref{equ8}) after including the Gaussian white noise as below,
	\begin{eqnarray}
	\dot{x}_{1}&=&-x_{1}+a_{1}y_{1}+s_{11}x_{1}+s_{12}x_{2}, \nonumber \\
	\dot{x}_{2}&=&-x_{2}+s_{21}x_{1}+s_{22}x_{2}+f_{1} \sin (\omega_{1} t)+f_{2} \sin (\omega_{2} t)+D\xi(t).~~~~~~~ 
	\label{equ11}
	\end{eqnarray}

	\begin{figure}[h!]
	\centering
	\includegraphics[width=0.8\columnwidth]{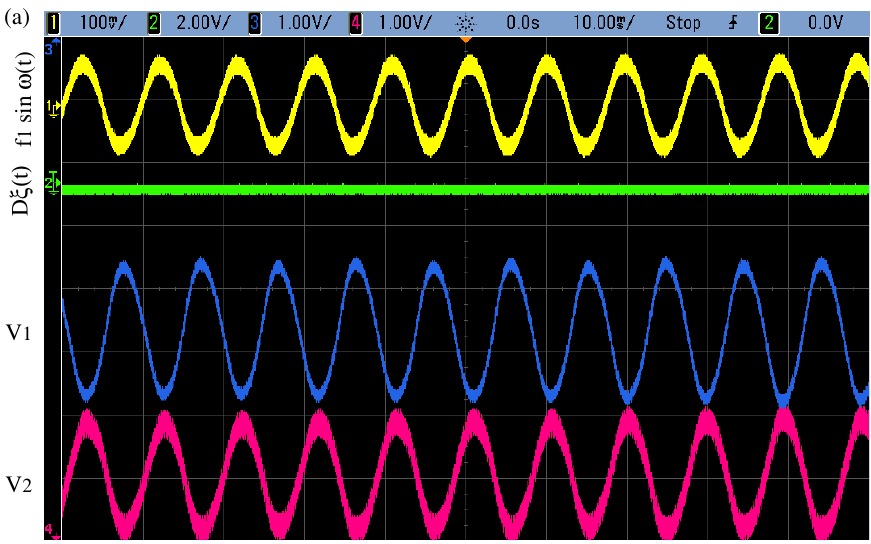}
	\includegraphics[width=0.8\columnwidth]{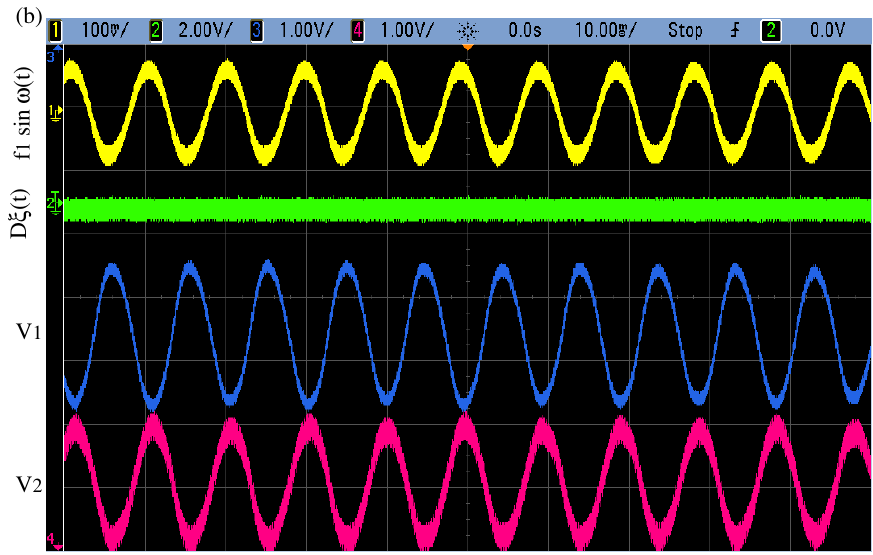}
	\includegraphics[width=0.8\columnwidth]{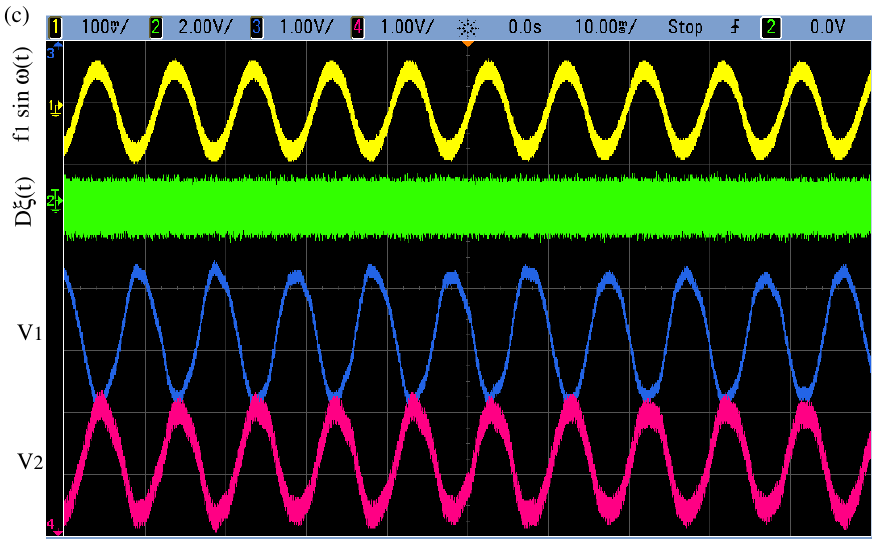}
	\includegraphics[width=0.8\columnwidth]{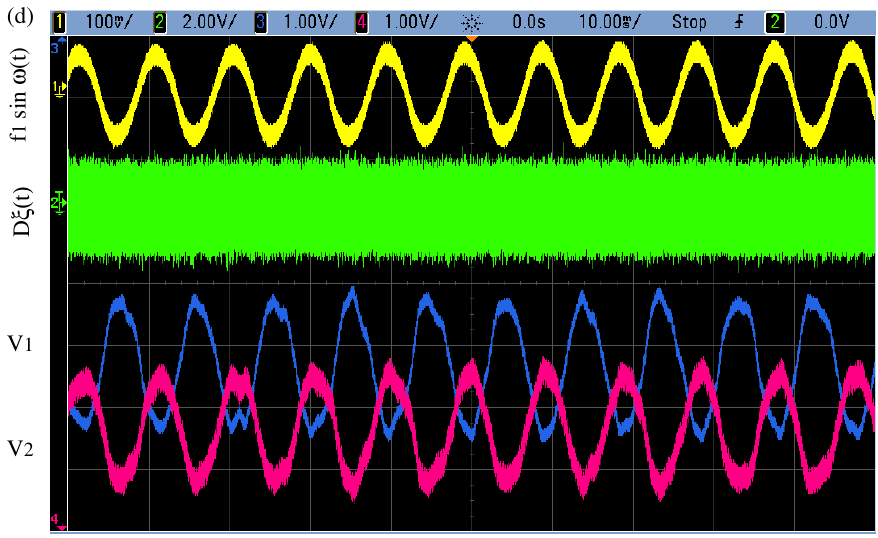}
	\caption{Realization of the signal detection in experimental electronic circuits: Panels (a)-(d) correspond to different values of $ D $ with fixed $ f_{1}=100$ mV and $ f_{2}=1$ V. Panel (a) represents $ D=0 $ V, panel (b) represents $ D=500$ mV, panel (c) represents $ D=1.5$ V, and panel (d) represents $ D=2.6 $ V. Every panel is having four subfigures namely  low amplitude $ f_{1} \sin \omega_{1}(t) $, Gaussian white noise $ D\xi(t) $, and the experimental outputs $ v_{1} $ and $ v_{2} $.}
	\label{fig9}
	\end{figure}

	In the above equation Eq.(\ref{equ11}) $ \xi(t) $ represents the Gaussian white noise and D is its strength. We fix the forcing parameters as $ f_1=0.25$ and $ f_2=25.0 $ and vary the noise strength $ D $. The numerically calculated time series outputs from  Eq.(\ref{equ11}) is shown in Fig.\ref{fig8}. Every panel in Fig.\ref{fig8} contains four figures, which correspond to the following: the low input signal $ f_1 $, noise $ D $, and the outputs $ x_1 $ and $ x_2 $. Let us consider the low input signal with $ f_1 = 0.25 $, a second forcing signal $ f_2 = 25.0 $, and noise $ D = 0.0 $. Fig.\ref{fig8}(a) depicts the corresponding outputs $x_1$ and $x_2$. In the form of inverse and similar signals, the two outputs, $x_1$ and $x_2$, continuously predict the frequency of the input signal, $f_1$. When the value of $ D $ is increased further to $D=0.1$, the outputs $x_1$ and $x_2$ continue to take the form of the input behavior [see Fig.\ref{fig8}(b)]. As the $D$ value is increased further to $D=0.5$, the outputs $x_1$ and $x_2$ gradually lose their actual input behavior [see Fig.\ref{fig8}(c)]. The output signals $ x_1 $ and $ x_2 $ then randomly oscillate and do not match with the low input signal $f_1$ as the value of $D$ increases to $D=1.0$ and beyond [see Fig.\ref{fig8}(d)]. 

	Next, we include the noise in the experimental SC-CNN based MLC circuit (Fig.\ref{fig2}(b)) in series with signal $ f_{1} $ and $ f_{2} $. By measuring the voltages $v_1$ and $v_2$ across the capacitors $C_1$ and $C_2$, respectively, the realization of experimental output and changes in the dynamics of the circuit under the effect of noise are obtained. In our experimental investigation, we are using an \textit{Agilent} function generator (33220A) and it has a noise signal option to generate the noise signal. The noise bandwidth of this function generator has been fixed at 9 MHz, which is typical. In our study, we vary the amplitude $ D \xi(t) $ of the noise signal (see Fig.\ref{fig9}(a-d) second channel). Further, we fix the forcing parameters as $ f_{1}=100$ mV  and $ f_{2}=1$ V and vary the noise parameter $D$ from $ 0 $ V to $ 2.6$ V. Fig.\ref{fig9}(a) depicts the low input signal $ f_1 $, the noise input signal $ D=0.0$ V, and two output signals $ v_1 $ and $ v_2 $ in this case. When no noise is present, $D=0.0$ V, the first and second inputs, $f_1$ and $f_2$, are set to $ 100 $ mV and $ 1 $ V, respectively. The frequencies of the corresponding outputs $ v_{1} $ and $ v_{2} $ exactly predict that of the lower input signal $ f_1 $ (as shown in Fig.\ref{fig9}(a)). As $D$ is increased to $500$ mV, the lower input signal is continued to be observed in the output signals (Fig.\ref{fig9}(b)). As the value $D$ is increased further to $ D=1.5 $~V(see Fig.\ref{fig9}(c)), the output signals gradually miss matching the input signals. When increasing the strength $ D $ to $ D = 2.6 $~V (see Fig.\ref{fig9}(d)), the system output signals are too weak to predict the low-frequency input signal. On further increasing the strength of the noise signal, one finds that the system no longer has the ability to mimic the input signal.

\begin{figure}[!ht]
\centering
\includegraphics[width=0.7\linewidth]{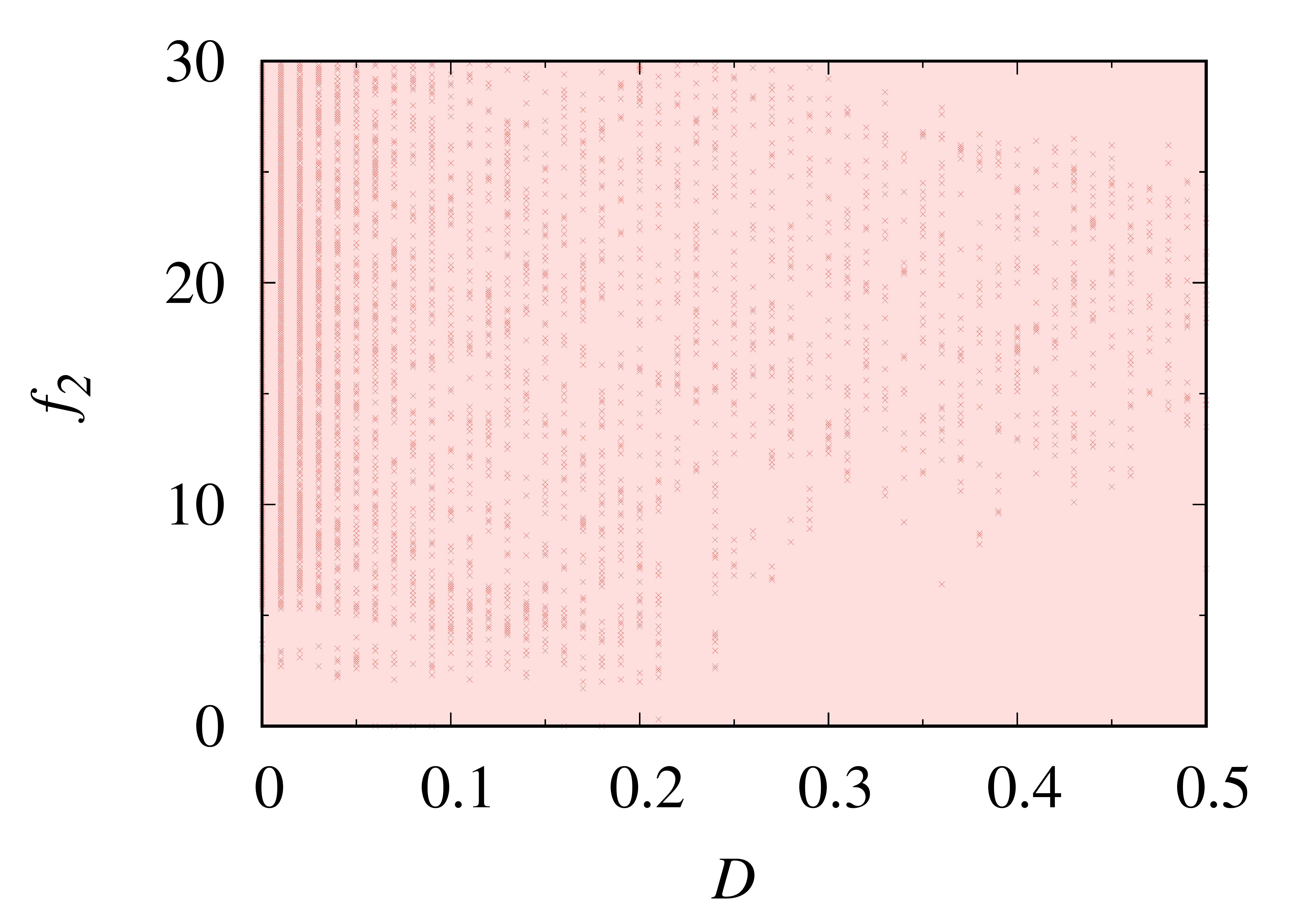}
\caption{The maximum response amplitude Q depicted in a two-parameter phase diagram for noise strength $ D $ vs second force $ f_2 $ for fixed parameters $ f_{1}=0.25 $, $\omega_1 = 0.75$ and $\omega_2 = 3.5$.}
\label{fig10}	
\end{figure}

	Fig.\ref{fig10} shows the probability of detecting or imitating the input signal after taking the noise strength $ (D) $ into account. The area to detect/mimic the weak input signal is indicated by the red dots. With strength $ D $, we determine the probability P(signal) of receiving the same input structure for various noise intensities. P(signal), in its simplest form, represents the proportion of total successful runs to total runs. P(logic) is given the value "1" if the system displays the intended input signal in response output; otherwise, it is handled as "0". A sampling of 2000 runs of the given input set is used to calculate P(signal) for the system (\ref{equ11}), and this process is repeated 1000 times. We use a low $ D $ value for the output signal in order to recognize or mimic the low input signal $ f_1 $. The noise progressively ceases matching the weak input signal $ f_1 $ as the noise level increases. Hence, even in the presence of noise coming from electronic components or any other external variables, the system's ability to recognize or replicate the low input signal is conclusively proven. 

	\section{Conclusions} 
	\label{sec7}

	In summary, we have demonstrated the existence of vibrational resonance in a nonlinear, nonautonomous SC-CNN-based MLC circuit. We have explained how the amplitude of the driving signal interacts with the state of the system through analytical, numerical and experimental studies. Particularly, we used two different forces with widely varying frequencies that contribute to generating vibrational resonance in the SC-CNN based MLC circuit. These forces work together to extend the region of resonance at specific system parameter values. Apart from this, we have also demonstrated the idea of detecting low input signals and obtaining enhanced output signals, which is one of the significant roles played by vibrational resonance in nonlinear systems. Besides these, we have also detected different low-input signals, such as square and sawtooth waves, in the same circuit without altering the system parameter values. We have also confirmed the robustness of detecting and enhancing the signal even after adding external Gaussian white noise.

	\section*{Acknowledgment}

	The authors wish to thank Dr. P. R. Venkatesh for providing support in the analytical calculations. A.V. thanks the DST-FIST for funding research projects via Grant No. SR/FST/College-2018-372 (C). M.L. wishes to acknowledge the DST-SERB National Science Chair program for funding under Grant No. NSC/2020/000029 in which PA and MS, respectively, are supported by a Project Associate and a Research Associate.

	\section*{Data availability statement}

	Data will be made available on request.

	\appendix

\section{Analytical evaluation of the response amplitude}
\label{sec_a}

Now the system (\ref{equ2}) can be explicitly integrated in terms of elementary functions in each of the three regions $ D_{0} $, $ D_{+} $ and $ D_{-} $ ($ |x| \le 1, x>1  $ and $ x<-1 $) and matched across the boundaries to obtain the full solution as shown below \cite{venkatesh2016vibrational}.

It is found that in each one of the regions $ D_{0} $, $ D_{+} $ and $ D_{-} $ the driving system (\ref{equ2})  can be represented as a single second order inhomogeneous differential equation for the variable $ y(t) $, 
 \begin{eqnarray}
	\ddot{y} & +(\beta+\beta v + \mu)\dot{y}+(\beta + \mu \beta v + \beta \mu)y = \bigtriangleup + \mu f_1 \sin (\omega_1t) \nonumber \\
	&  + \omega_1 f_1 \cos (\omega_1t)+ \mu f_2 \sin (\omega_2 t) + \omega_2 f_2 \cos (\omega_2 t),
	\label{a1}
\end{eqnarray}
where 

(i)  $\mu = a,$ $\bigtriangleup = 0$ in region $D_0$, and

(ii) $\mu = b,$ $\bigtriangleup = \pm\beta (a-b)$ in region $D_{\pm}$.

The general solution of the system \ref{a1} can be written as
	\begin{eqnarray}
			y(t) &=& C_{0}^{1},_{\pm} exp(\alpha_{1}t) + C_{0}^{2},_{\pm} exp(\alpha_{2}t)+ E_{1}+E_{12} \sin (\omega_{1}t) \nonumber \\ &&+ E_{13}\cos (\omega_{1}t)+E_{22} \sin (\omega_{2}t) + E_{23}\cos (\omega_{2}t),
			\label{a2}
	\end{eqnarray}
where $ C_{0}^{1},_{\pm}, C_{0}^{2},_{\pm} $ are integration constants in the appropriate regions $ D_{0} $, $ D_{\pm} $, and 
\begin{eqnarray*}
		\alpha_{1,2} &=& (-A \pm \sqrt{A^2 - 4B})/2,  \\
		A  &=& \beta + \beta v + \mu, ~~ B =\beta + \mu \beta v + \beta \mu, \\
		E_1  &=& 0~ in~ region~ D_0, ~~	E_1 = \bigtriangleup / B~ in ~region ~D_\pm,\\
		E_{12}  &=& \frac{[f_1 \omega_1^2  (A-\mu) + \mu f_1 B]}{[A^2 \omega_1 ^2 + (B - \omega_1 ^2)^2]}, ~~ E_{13}  =\frac{f_1 \omega_1[B - \omega_1 ^2 - \mu A]}{[A^2 \omega_1 ^2 + (B - \omega_1 ^2)^2]},\\
		E_{22}  &=& \frac{[f_2 \omega_2^2  (A-\mu) + \mu f_2 B]}{[A^2 \omega_2 ^2 + (B - \omega_2 ^2)^2]}, ~~ E_{23}  =\frac{f_{2} \omega_2[B - \omega_2 ^2 - \mu A]}{[A^2 \omega_2 ^2 + (B - \omega_2 ^2)^2]}.		
\end{eqnarray*}
Knowing $ y(t) $, we can obtain $ x(t) $ from (\ref{equ2}) as
\begin{eqnarray}
	x(t) = 1/\beta [-\dot{y}-\beta y (1+\nu) + f_{1} \sin (\omega_{1})t + f_{2} \sin (\omega_{2}) t].~~~~
	\label{a3}
\end{eqnarray}
Substituting for $ y $ and $ \dot{y} $ from Eq.(\ref{a2}), the form of $ x(t) $ is found to be
\begin{eqnarray}
	x(t) = 1/\beta \big[-C_{0}^{1},_{\pm} (\alpha_{1}+\sigma) exp(\alpha_{1}t) - C_{0}^{2},_{\pm} (\alpha_{2}+\sigma)exp(\alpha_{2}t) \nonumber \\ +(E_{12}\omega_{1}+E_{13}\sigma) \cos (\omega_{1} t) + (f_{1}-E_{12}\sigma+E_{13}\omega_{1}) \sin (\omega_{1} t) \nonumber \\ +(E_{22}\omega_{2}+E_{23}\sigma) \cos (\omega_{2} t) + (f_{2}-E_{22}\sigma+E_{23}\omega_{2}) \sin (\omega_{2} t) \nonumber \\ - E_{1}\sigma \big].~~~~~~
	\label{a4}
\end{eqnarray}
where $\sigma = \beta (1+v)$ and the value of the integration constants, namely $C_{0,\pm}^1$ and $C_{0,\pm}^2$ are obtained by substituting the initial condition at $t=t_0$, $x(t_0)=x_0$ and $y(t_0)=y_0$ in Eqs. (\ref{a3}) and (\ref{a4}). On solving further, the explicit form of the integration constants are found as
\begin{eqnarray}
		C_{0,\pm}^1 & = &\exp (-\alpha_1 t_0)/ (\alpha_1 - \alpha_2) [-\beta x_0 + [E_{12} \omega_1 + E_{13} (\alpha_2 \nonumber \\ && + 2 \sigma)]  \cos (\omega_1 t_0)  + [f_1 + E_{12} \alpha_2 + E_{13} \omega_1] \sin (\omega_1 t_0) \nonumber \\ && + [E_{22 \omega_2} + E_{23} (\alpha_2 + 2 \sigma)] \cos (\omega_2 t_0 )  
		 + [f_2 + E_{22} \alpha_2 \nonumber \\ && + E_{23} \omega_2] \sin (\omega_2 t_0) - E_1 \alpha_2 - y_0 (\alpha_2 + \sigma)],
		 \label{a5}
\end{eqnarray} 
	\begin{eqnarray}
			C_{0,\pm}^2 & = & \exp (-\alpha_1 t_0)/ (\alpha_2 - \alpha_1) [-\beta x_0 + [E_{12} \omega_1 + E_{13} (\alpha_1 \nonumber \\ && + 2 \sigma)]   \cos (\omega_1 t_0) + [f_1 + E_{12} \alpha_1 + E_{13} \omega_1] \sin (\omega_1 t_0) \nonumber \\ && + [E_{22 \omega_2} + E_{23} (\alpha_1 + 2 \sigma)] \cos (\omega_2 t_0 ) + [f_2 + E_{22} \alpha_1  \nonumber \\ && + E_{23} \omega_2] \sin (\omega_2 t_0) - E_1 \alpha_1 - y_0 (\alpha_1 + \sigma)].
			\label{a6}
	\end{eqnarray}

The response of the system is than calculated from the sine  and cosine components, $ Q_{s} $ and $ Q_{c} $, respectively of the output signal $ x(t) $. 

Using Eq.(\ref{a4}) the sine and cosine constituents of the output signal are given by
\begin{eqnarray}
	Q_{s} = \frac{2}{nT} \int_{0}^{nT} x(t) \sin (\omega_{1} t) dt, \nonumber \\
	Q_{c} = \frac{2}{nT} \int_{0}^{nT} x(t) \cos (\omega_{1} t) dt.
	\label{a7}
\end{eqnarray}	
where $ T=2\pi/\omega_{1} $ and n is a positive integer.\\
Then, we finally find the dependence on $ f_{2} $ of the response amplitude $ Q_{ana} $ as
\begin{eqnarray}
	Q_{ana} = \frac{\sqrt{\rule{0pt}{2ex} Q_{s}^2+Q_{c}^2}}{f_{1}}, 
	\label{a8}
\end{eqnarray}

In order to evaluate the analytical expression for response amplitude, one can rewrite the Eq. (\ref{a4}) as 
\begin{eqnarray}
		x(t) & = 1/ \beta [- C_{0,\pm}^1 (\alpha_1 + \sigma) \exp (\alpha_1 t) - C_{0,\pm}^2 (\alpha_2 + \sigma) \exp (\alpha_2 t) \nonumber \\
		& + S \cos (\omega_1 t- \phi_1)  + S' \cos (\omega_2 t - \phi_2) - E_1 \sigma], 
		\label{a9}
\end{eqnarray} 
where
\begin{eqnarray}
		S^2  = (E_{13}^2 + E_{12}^2) (\omega_1^2 + \sigma^2) + f_1^2 + 2f_1 \omega_1 E_{13} - 2f_1 \sigma E_{12}, ~~~~~
		\label{a10}
\end{eqnarray}
\begin{eqnarray}
		S^{'2}  = (E_{23}^2 + E_{22}^2) (\omega_2^2 + \sigma^2) + f_2^2 + 2f_2 \omega_2 E_{23} - 2f_2 \sigma E_{22},~~~~~
		\label{a11}
\end{eqnarray}
\begin{eqnarray}
		\phi_1  = \tan^{-1} \frac{f_1 - E_{12} \sigma + E_{13} \omega_1}{E_{12} \omega_1 + E_{13} \sigma},
		\label{a12}
\end{eqnarray} 	
\begin{eqnarray}
		\phi_2 = \tan^{-1} \frac{f_2 - E_{22} \sigma + E_{23} \omega_2}{E_{22} \omega_2 + E_{23} \sigma}.
		\label{a13}
\end{eqnarray} 
By substituting Eq.(\ref{a9}) in Eq. (\ref{a7}) one can get
\begin{eqnarray}
		Q_s^{0,\pm} = \frac{2}{nT\beta} \int_{0}^{nT} [-C_{0,\pm}^1 (\alpha_1 + \sigma) \exp (\alpha_1 t) - C_{0,\pm}^2 (\alpha_2 + \sigma) \nonumber \\ \exp (\alpha_2 t) 
		 + S \cos (\omega_1 t - \phi_1) + S' \cos (\omega_2 t - \phi_2) - E_1 \sigma] \nonumber \\ sin (\omega_1 t ) dt, ~~~~~~~~
		 \label{a14}
\end{eqnarray} 
\begin{eqnarray}
		Q_c^{0,\pm} = \frac{2}{nT\beta} \int_{0}^{nT} [-C_{0,\pm}^1 (\alpha_1 + \sigma) \exp (\alpha_1 t) - C_{0,\pm}^2 (\alpha_2 + \sigma)  \nonumber \\ \exp (\alpha_2 t)
		 + S \cos (\omega_1 t - \phi_1) + S' \cos (\omega_2 t - \phi_2) - E_1 \sigma]  \nonumber \\ cos (\omega_1 t ) dt. ~~~~~~~~
		 \label{a15}
\end{eqnarray}
During the time interval $ nT $, the value of $ x(t) $ flips between various regions namely, $ D_{0} $, $ D_{+} $, and $ D_{-} $, and the intrusion times of $ x(t) $ in each of the regions are identified separately. Then the sine and cosine components of $ x(t) $ in each of the components of sine $ (Q_{s}^{0,\pm}) $ and cosine $ (Q_{c}^{0,\pm}) $ are added successively depending on their dwelling in the three different regions $ D_{0} $, $ D_{\pm} $. Finally the response amplitude $ Q $ is calculated as
\begin{eqnarray}
		Q_{ana} =  \frac{\sqrt{(\sum Q_s^{0,\pm})^2 + (\sum Q_c^{0,\pm})^2}}{f_1},
		\label{a16}
\end{eqnarray}
to obtain the analytical response amplitude.

\end{document}